# JANC: A cost-effective, differentiable compressible reacting flow solver featured with JAX-based adaptive mesh refinement

Haocheng Wen[1,*], Faxuan Luo[1,*], Sheng Xu, Bing Wang[*]

Tsinghua University, School of Aerospace Engineering, Beijing, China, 100084

**Abstract**: The compressible reacting flow numerical solver is an essential tool in the study of combustion, energy disciplines, as well as in the design of industrial power and propulsion devices. We have established the first JAX-based block-structured adaptive mesh refinement (AMR) framework, called JAX-AMR, and then developed a fully-differentiable solver for compressible reacting flows, named JANC. JANC is implemented in Python and features automatic differentiation capabilities, enabling an efficient integration of the solver with machine learning. Furthermore, benefited by multiple acceleration features such as XLA-powered JIT compilation, GPU/TPU computing, parallel computing, and AMR, the computational efficiency of JANC has been significantly improved. In a comparative test of a two-dimensional detonation tube case, the computational cost of the JANC core solver, running on a single A100 GPU, was reduced to 1% of that of OpenFOAM, which was parallelized across 384 CPU cores. When the AMR method is enabled for both solvers, JANC's computational cost can be reduced to 1-2% of that of OpenFOAM. The core solver of JANC has also been tested for parallel computation on a 4-card A100 setup, demonstrating its convenient and efficient parallel computing capability. JANC also shows strong compatibility with machine learning by combining adjoint optimization to make the whole dynamic trajectory efficiently differentiable. JANC provides a new generation of high-performance, cost-effective, and high-precision solver framework for large-scale numerical simulations of compressible reacting flows and related machine learning research. Now, the source codes have been available under the MIT license at https://github.com/JA4S/JAX-AMR and https://github.com/JA4S/JANC.

**Key words**: JAX, adaptive mesh refinement, open source, compressible reactive flow, combustion, computational fluid dynamics

## 1. Introduction

The numerical simulation of compressible reactive flow is a critical issue and challenge in computational fluid dynamics (CFD), with applications spanning combustion systems [1], energy industry [2], aerospace propulsion [3], and astrophysical phenomena [4]. The flow phenomena in these fields are characterized by complex multi-physics and multi-scale interactions, including shock waves, turbulence, and chemical reactions, etc. Accurate simulations of these flow phenomena rely on reliable numerical methods, precise physical models, and high spatial resolution. Traditional compressible reactive flow solvers have faced numerous challenging problems when attempting to further enhance computational accuracy and efficiency, including developing numerical schemes suitable for multi-scale and multi-type discontinuities, accurate turbulence models and chemical reaction mechanisms, and the need to further improve spatial resolution.

In recent years, the introduction of machine learning (ML) frameworks has provided

[1] Both authors contributed equally.

[*] Corresponding authors. Email: haochengwenson@126.com (HC Wen), luofx23@mails.tsinghua.edu.cn (FX Luo), wbing@mail.tsinghua.edu.cn (B Wang)

transformative tools for addressing these challenges, opening new avenues for the development of CFD solvers and demonstrating significant potential for growth. For example, the WENO3-NN scheme based on neural networks can adapt better to different resolutions and flow problem solutions [5]. Data-driven turbulence surrogate models enhance the applicability and solution accuracy of turbulence models [6]. Chemical reaction mechanisms derived from neural network training can reduce computational overhead while ensuring the predication accuracy of chemical reaction [7]. High-resolution reconstruction of low-resolution flow fields can also be achieved through ML approaches, such as CNN and DSC/MS models [8]. Additionally, various large language models (LLMs) for fluid dynamics have been utilized to accelerate CFD computations [9].

However, traditional CFD solvers still face several challenges when integrated directly with ML frameworks. These include poor generalization performance [10] and significant error accumulation during time integration, which often leads to numerical instability [11]. By embedding the numerical computation process directly into the optimization loop of machine learning, allowing real-time interaction between the optimizer and the solver, these issues can be significantly mitigated [11]. Such solvers combine traditional numerical methods with Automatic Differentiation (AD) techniques, making the entire computational pipeline differentiable. This enables gradient-based optimization and learning, supporting a wide range of data-driven tasks such as reduced-order modeling of flow fields [12] and flow field reconstruction [13].Currently, various differentiable numerical computing frameworks have been proposed, including TensorFlow [14], PyTorch [15], JAX [16], etc. Additionally, traditional CFD solvers are primarily developed for CPU computing, making them difficult to adapt to next-generation computing architectures like GPUs and TPUs. This not only limits the further enhancement of the solver's computational efficiency but also significantly increases the difficulty of combining the solvers with the numerous deep learning models currently developed on GPU architectures.

The new generation of differentiable CFD solvers based on JAX offers new possibilities. JAX is a Python library designed for accelerator-oriented array computation, optimized for GPUs and TPUs, featuring automatic differentiation, XLA-powered JIT (Just-In-Time) compilation, and hardware parallel optimization, which enable high-performance scientific computing and machine learning research [16]. At present, a few of JAX-based solvers for different fields have been released, such as JAX-MD [17], JAX-FEM [18], JAX-FLUIDS [19], etc. Among them, JAX-FLUIDS is the first fully differentiable CFD solver for compressible two-phase flows. These differentiable solvers have already demonstrated unique advantages when integrated with ML, including enabling end-to-end optimization [19].

Nevertheless, two critical areas are still lacking within the community of JAX-based solvers. First, there is currently no JAX-based solver developed specifically for compressible reacting flow, which fails to meet the demands of related fields. More importantly, there is no adaptive mesh refinement (AMR) framework compatible with JAX, which greatly increases the cost and difficulty of conducting large-scale, high spatial resolution numerical simulations. The main difficulty of JAX-based AMR lies in the restriction that JIT in JAX requires the program computation graph to be static; that is, the shapes of all arrays must be known and fixed at compile time. In existing AMR methods, however, array shapes are typically dynamic, including quadtree/octree-based methods (such as OpenFOAM [20]) and block-structured methods (such as AMReX [21], AMROC [22]). This dynamic shape input results in the function being re-JITed upon each call, which incurs an unacceptable compilation overhead.

The primary contribution of our work is the establishment of the first fully JIT-compatible AMR framework, JAX-AMR, and the development of a compressible reacting flow solver, named JANC, based on this framework. JANC has undergone benchmark case testing and has demonstrated impressive improvements in computational efficiency compared to traditional open-source CFD solvers like OpenFOAM [23] in benchmark case comparisons. Furthermore, by integrating JANC's solver with the adjoint optimization algorithm [24], we demonstrate that the complete numerical solution pipeline in JANC can perform differentiable computation with low memory and high computational efficiency. This capability enables seamless integration with a wide variety of downstream machine learning tasks. JANC provides a new generation of numerical simulation tools for research in the field of compressible reacting flow.

The remainder of this paper is organized as follows: Section 2 briefly introduces the basic features of JANC. Section 3 introduces the implementation strategy of JAX-AMR. Section 4 elaborates the implementation details of the compressible reacting flow solver. Section 5 presents the validation cases and comparative analysis on computation performance. The differentiable capabilities and adjoint optimization methods are discussed in Section 6, followed by conclusions and perspectives in Section 7.

## 2. Basic features of JANC

JANC, as the abbreviation for "JAX-AMR & Combustion", is a fully-differentiable compressible reacting flow solver based on JAX. The basic features of JANC are as follows:

- Implementation of adaptive mesh refinement (AMR) based on JAX, namely JAX-AMR, providing a feasible AMR framework for large-scale fully-differentiable computation.
- Adoption of Cartesian structured grid, dimensionless equations, high-order finite difference method, point-implicit chemical source advancing in the solver.
- Inheriting the basic features of JAX, including fully-differentiable, compatible with CPUs/GPUs/TPUs computation, and convenient parallel management.
- Programmed by Python, allowing rapid and efficient prototyping of projects.

Benefiting by the JIT, AMR and GPUs acceleration, it will be demonstrated later that, the computational efficiency of JANC is significantly improved compared to mainstream open-source solvers such as OpenFOAM. Additionally, JANC enables differentiable computation with low memory usage and high computational speed, allowing seamless integration with a wide range of downstream machine learning tasks.

## 3. JAX-AMR: JAX-based adaptive mesh refinement framework

We propose an adaptive mesh refinement framework based on dynamically updated multi-layer blocks with fixed positions and fixed shapes, referred to as JAX-AMR. This framework is fully compatible with JIT and vectorized operations (`jax.vmap`). Furthermore, similar to all block-structured AMR frameworks, such as AMReX [21], the load balancing during parallel computation of blocks at different levels will be easily achievable in the future version.

This section first introduces the basic concept and implementation strategy of blocks within JAX-AMR, followed by the implementation details of the program and the fundamental ways to integrate JAX-AMR with the CFD solver.

## *3.1 Basic concepts and strategy of block*

The multi-layer block structure adopted in JAX-AMR is illustrated in Figure 1. For clarity, the set of blocks in the *i*-th refinement layer is denoted as $\mathrm{B}_i$,

$$\mathrm{B}_i = \{\mathrm{b}_{i,j} \mid j = [0, N_{i,\max} - 1]\}$$

where $\mathrm{b}_{i,j}$ represents the blocks contained in $\mathrm{B}_i$, and $N_{i,\max}$ is the maximum number of blocks in each $\mathrm{B}_i$.

Note that, in JAX-AMR, to ensure compatibility with JIT and vectorized operations, all $\mathrm{b}_{i,j}$ within the same layer must have the same size and shape. Furthermore, to avoid the compilation overhead associated with frequent JIT, $N_{i,\max}$ also needs to be predeclared as a fixed value in the current version. This predeclared value may need to be much greater than the number of valid blocks $N_{i,\text{valid}}$ that are actually refined within the computational domain, in order to ensure sufficient redundancy for the potential addition of valid blocks during computation, which leads to unnecessary memory and computational overhead. To address this issue, JAX-AMR temporarily employs a strategy of dynamically updating $N_{i,\max}$ at irregular intervals, which will be introduced in more detail later.

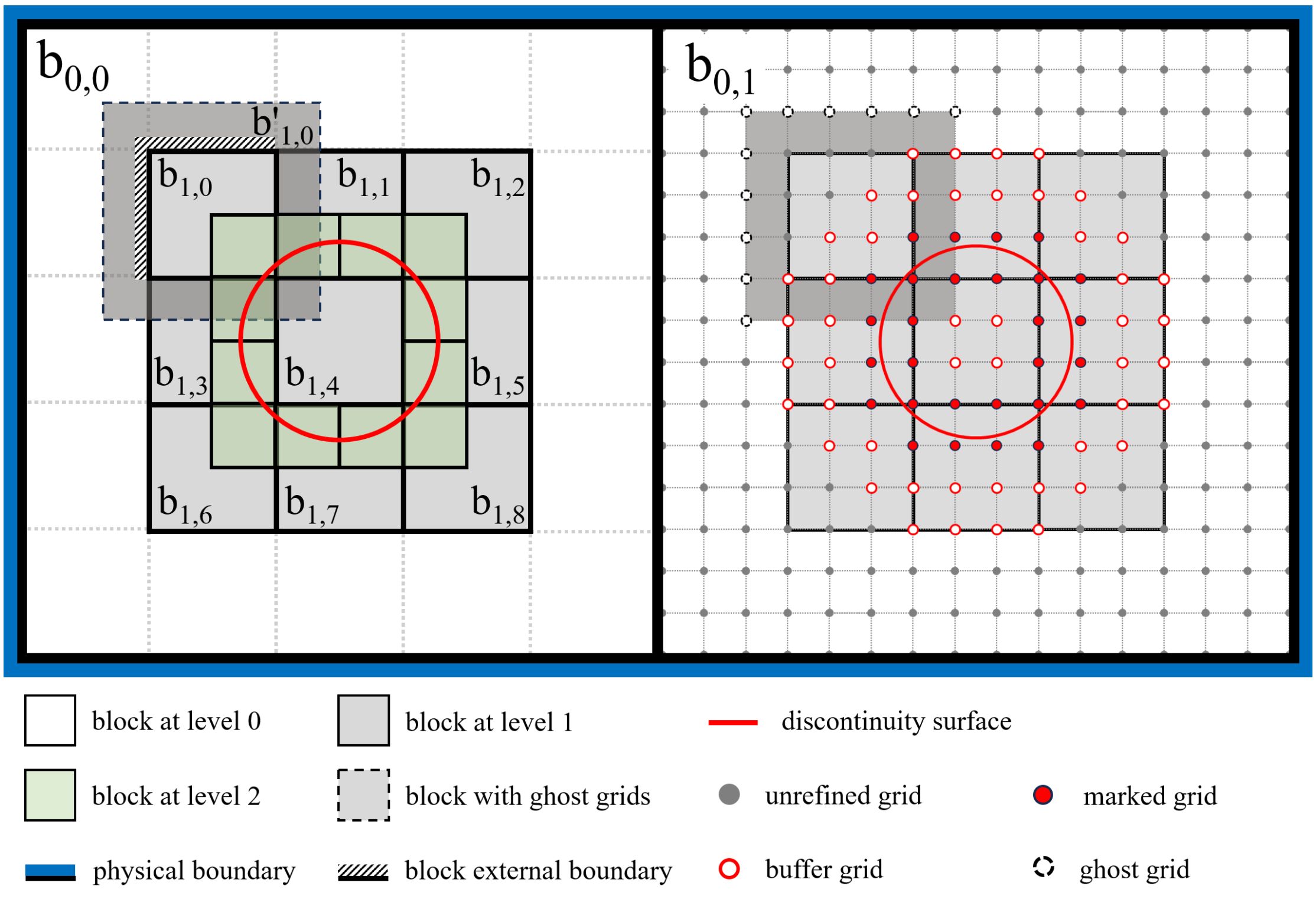

**Figure 1.** Schematic diagram of multi-layer blocks and the partitioning and refinement strategies in JAX-AMR.

### *3.1.1 Partitioning strategy of blocks*

Here is the partitioning strategy for blocks in each refinement layer. First, we have the base layer $\mathrm{B}_0$, which corresponds to the area within the blue box in Figure 1, and represents the complete computational domain. In Figure 1, $\mathrm{B}_0$ is evenly divided into two blocks, namely $\mathrm{b}_{0,0}$ and $\mathrm{b}_{0,1}$. In JAX-AMR, $\mathrm{B}_0$ is primarily set up for parallel computation at the base layer. Since $\mathrm{B}_0$ does not involve refinement decisions or dynamic changes, the maximum block number $N_{0,\max}$ can be predicted and determined before compilation.

Building on $B_0$, the first refinement layer $B_1$ can be partitioned. First, $b_{0,0}$ is divided into several equally shaped subregions (5×5 subregions in Figure 1) with fixed positions. Next, the blocks that need refinement are identified, which are represented by $b_{1,0}$ to $b_{1,8}$ in Figure 1. Note that, due to the requirement that the declared $N_{1,\max} > 9$, the redundant blocks are only declared in memory and does not correspond to actual flow field regions.

The above process is repeated, continuing to establish the second refinement layer $B_2$ based on $B_1$, and then deeper refinement layers.

### *3.1.2 Refinement strategy of blocks*

The refinement strategy of blocks in JAX-AMR is illustrated in the block $b_{0,1}$ on the right side of Figure 1. First, within each layer of grids, specific refinement criteria are applied to identify the grids that need refinement, namely masked grids. For example, the refined grids are chosen as the grids on either side of a discontinuity in Figure 1. Next, several additional grids are extended outward from the masked grids, namely buffer grids. The purpose of these buffe grids is to ensure that the structures that require refinement do not move outside the refinement region before the AMR update. The number of buffe grids must be no less than the propagation distance of the disturbance in the unit AMR update step. Finally, the subregions containing the masked grids and the buffe grids are designated as blocks that are to be refined.

### *3.1.3 Basic attributes of blocks*

Each block in JAX-AMR contains two basic attributes: **data** and **info**, represented as follows:

$$b_{i,j} = \{\text{data, info}\}$$

The **data** attribute is a multi-dimensional JAX array that stores the physical quantities at the grids within the block. For a two-dimensional problem, the shape of **data** is $[n_U, n_x, n_y]$, where $n_U$ is the number of physical quantities, and $n_x$ and $n_y$ represent the number of grids in the $x$- and $y$- directions, respectively.

The **info** attribute records relevant information for the corresponding block:

$$\text{info} = \{\text{index, glob_index, neighbor_index}\}$$

The specific elements within **info** include:

- **index**: The relative index of $b_{i,j}$ in the previous layer $B_{i-1}$, represented as a three-dimensional coordinate $(k, r, c)$, where $k$ indicates that $b_{i,j}$ is the $k$-th block $b_{i-1,k}$ in the previous layer, and $(r, c)$ indicates that $b_{i,j}$ is the subregion located at $r$-th row and $c$-th column in $b_{i-1,k}$. For example, the relative index of $b_{1,5}$ in Figure 1 is (0, 1, 2), where the starting values for $k$, $r$, $c$ are all 0.
- **glob_index**: The global index of $b_{i,j}$ across all refinement layers, represented as a nested multi-dimensional coordinate $(a, r, c)_{i,j}$, where $a_{i,j} = (a, r, c)_{i-1,k}$. For example, the global index of $b_{1,5}$ in Figure 1 is ((0, 0), 1, 2).
- **neighbor_index**: The index of neighboring blocks for $b_{i,j}$ in the current layer $B_i$, represented as a four-dimensional coordinate $(n, s, w, e)$, where $n$, $s$, $w$, $e$ correspond to the sequence numbers of the neighboring blocks to the north, south, west, and east, respectively, within $B_i$. If there are no neighboring blocks in a certain direction, the corresponding coordinate is marked as −1. For instance, the neighbor index of $b_{1,5}$ in

Figure 1 is (2, 8, 4, −1).

Each block also has a **ghost block** and three types of boundaries.

The ghost block is generated by extending several grids beyond the outer layer of the block, as shown in $b'_{1,0}$ in Figure 1. The number of extended grid points, denoted as $n_{\text{ghost}}$, must satisfy $n_{\text{ghost}} \geq n_{\text{template}}$, where $n_{\text{template}}$ refers to the template nodes number on one side of the numerical scheme. This condition is designed to ensure that when the refined ghost block is used as input for the solver, there are available template nodes for the computation of the boundary grids of the original block.

The three types of boundaries are as follows:

- **interior boundary**: The shared boundary between adjacent blocks. For example, the southern and eastern boundaries of $b_{1,0}$ in Figure 1.
- **external boundary**: The boundary on the side without neighboring blocks, where the exterior consists of the grids from the previous refinement layer. For instance, the northern and western boundaries of $b_{1,0}$ in Figure 1.
- **physical boundary**: The external boundary of the base layer $B_0$, which serves as the physical boundary of the entire computational domain. For example, the northern, southern, and western boundaries of $b_{0,0}$ in Figure 1.

Finally, as introduced at the beginning of this section, each block is distinguished by valid and invalid attributes in JAX-AMR.

- **valid block**: For $0 \leq j \leq N_{i,\text{ valid}} - 1$, the blocks $b_{i,j}$ are valid blocks, which correspond to the actual computational domain.
- **invalid block**: For $N_{i,\text{ valid}} \leq j \leq N_{i,\text{ max}} - 1$, the blocks $b_{i,j}$ are reserved as invalid blocks. These blocks do not correspond to any actual computational domain. Although their computed values do not affect the overall results, they will incur memory and computational overhead. The primary purpose of invalid blocks is to reserve space for valid blocks that may be added after an AMR update, avoiding for frequent calls to JIT.
- **NAN block**: Specifically, in JAX-AMR, the last block $b_{i,-1}$ in each layer $B_i$ is marked as a NAN block, with all values in its data and info set to NAN. This is to mark the values of ghost grids without neighbor block as invalid values NAN. Therefore, it is important to note that if the solver involves computing global extrema, averages, or similar operations, the influence of NAN values needs to be removed. For instance, the function `jnp.nanmax` should be used instead of `jnp.max`.

### *3.1.4 Basic operations of blocks*

The basic operations related to blocks in JAX-AMR mainly include refinement, coarsening, obtaining ghost blocks, solving block data, updating external boundaries, and updating blocks.

- **refinement**: The **data** attributes in $b_{i,j}$, which is partitioned based on layer $B_{i-1}$ ($i \geq 1$), is refined through interpolation. Currently, JAX-AMR employs a "one to four" method for refinement interpolation, assigning the same values of the coarse grids to the refined grids.
- **coarsening**: The **data** attributes in $b_{i,j}$ ($i \geq 1$) is coarsened through interpolation. JAX-AMR currently uses a "four to one" method for coarsening interpolation, where the arithmetic average of the values of the refined grids is assigned to the corresponding grids in $B_{i-1}$.
- **obtaining ghost block**: The neighbor blocks of $b_{i,j}$ are indexed through **neighbor_index**, and the ghost grids in $b'_{i,j}$ are assigned the values from the neighbor block's grids.
- **solving block data**: The **data** of ghost block is input in a vectorized form into the solver

for computation.

- **updating external boundary**: The values of several layers of grids near the external boundary in the solved $b_{i,j}$ are assigned the corresponding grid values from the unsolved $b_{i,j}$. This is done to ensure that the boundary values are synchronized with the data of layer $B_{i-1}$. The number of grids to be synchronized is given by $n_{boundary} = n_{template}$.
- **updating refinement layer**: The layer $B_i$ is updated based on the current physical quantities, while retaining the **data** attributes of blocks that have the same **glob_index** before and after the update.

Particularly, in the current version of JAX-AMR, a dynamic update strategy for $N_{i,\ max}$ is employed tentatively, as shown in Algorithm 1.

**if** $(N_{i,\ valid} + 1) > N_{i,\ max}$
| $N_{i,\ max} = 2\ N_{i,\ max}$
**elif** $(N_{i,\ valid} + 1) < 2.5\ N_{i,\ max}$
| $N_{i,\ max} = N_{i,\ max} / 2$
**end**

**Algorithm 1**: Dynamic update strategy for $N_{i,\ max}$ in current version of JAX-AMR

Through the above update strategy, it is possible to minimize the computational and memory overhead caused by a large number of redundant blocks while avoiding the frequent JIT compilation costs that come with updating $N_{i,\ max}$. However, the current updating strategy is only a simple version and still incurs some redundant computational overhead. In the future, a more intelligent strategy will be adopted to further reduce costs.

*3.2 Implementation details in JAX-AMR*

This section introduces the main algorithms involved in JAX-AMR as well as some specific programming details.

JAX-AMR is implemented using pure Python. Because JAX leans towards a functional programming style, the current version of JAX-AMR does not encapsulate AMR-related functions within classes; rather, all major functions are independent and compatible with JIT.

The program file structure of the current version of JAX-AMR is shown in Table 1.

**Table 1.** File structure of current version of JAX-AMR

```
JAX-AMR/src
    ├── jaxamr.py       # core functions for blocks operations
    ├── amrsolver.py    # time advance functions with AMR
    ├── amraux.py       # auxiliary functions, including the pre/post-process functions
JAX-AMR/examples
    ├── main.py         # example CFD program based on JAX-AMR
    ├── config.py       # configuration file for the example program
    ├── cfd_solver.py   # simple CFD solver for Euler equations
```

Algorithm 2 describes how to implement the main loop of a CFD program using functions from JAX-AMR in **main.py**. When entering the loop for the first time, it is necessary to initialize each refinement level, which involves calling the following functions in sequence:

- `get_refinement_grid_mask`: This function generates the refinement grid mask, `refinement_grid_mask`, based on the given refinement criteria. Users are allowed customizing the criteria within this function.
- `get_refinement_block_mask`: This function generates the blocks mask to be refined, `refinement_blk_mask`, based on `refinement_grid_mask`.
- `initialize_max_block_number`: This function estimates $N_{i,\max}$, the block number to be reserved in $B_i$ based on the number of valid blocks $N_{i,\text{valid}}$ in `refinement_blk_mask`.
- `get_refinement_block_info`: This function generates the **info** attributes for $b_{i,j}$ in $B_i$ based on `refinement_blk_mask` and stores them sequentially in a JAX Pytrees type variable called `blk_info`.
- `get_refinement_block_data`: This function generates the **data** attributes for $b_{i,j}$ in $B_i$ based on `refinement_blk_mask`, refines the data, and stores the results in a JAX Array type variable called `blk_data` with the shape $[N_{i,\max}, n_U, n_x, n_y]$.

Next, it is determined whether AMR should be updated. The AMR update shares several common functions with AMR initialization to update `blk_info` and `blk_data` for each level of $B_i$. The other functions called include:

- `update_max_block_number`: This function determines the current $N_{i,\text{valid}}$ based on the updated `refinement_blk_mask` and re-estimates and updates $N_{i,\max}$.
- `find_and_update_unaltered_block`: This function identifies the unaltered blocks based on the original and updated `blk_info`, keeping their original `blk_data`.

Subsequently, the `solver` function is used to solve and update `blk_data` for each level of $B_i$. Two time-advance strategies are available. The recommended strategy is the crossover advance strategy, as shown in Figure 2a, where the deepest refinement level is solved first: for every two steps taken in the current level's solution, one step is taken to update the previous level's solution. Note that when using this time-advance strategy, the time step for each level's solution is half that of the previous level. Alternatively, the synchronous advance strategy illustrated in Figure 2b can be used, where each level is solved and updated synchronously, and the time step for each level should match that of the deepest level.

Finally, the function `interpolate_fine_to_coarse` is employed to interpolate the `blk_data` of each level to the previous level and update the `blk_data` of the previous level.

```
for step do
    if initialize do    # initialize AMR blocks at level 1, 2, …
        get_refinement_grid_mask ()
        get_refinement_block_mask ()
        initialize_max_block_number ()
        get_refinement_block_info ()
        get_refinement_block_data ()
    elif update do    # update AMR blocks at level 1, 2, …
        get_refinement_grid_mask ()
        get_refinement_block_mask ()
        update_max_block_number ()
        get_refinement_block_info ()
```

```
        get_refinement_block_data ()
        find_and_update_unaltered_block ()
    end

    for 1, 2 do   # solve AMR blocks at level 0, 1, 2, …
        for 1, 2 do
            solver () # at level 2
        end
        solver ()  # at level 1
    end
    solver () # at level 0

    interpolate_fine_to_coarse () # at level …, 2, 1
end
```

**Algorithm 2**: Main loop of a CFD program conjunction with JAX-AMR

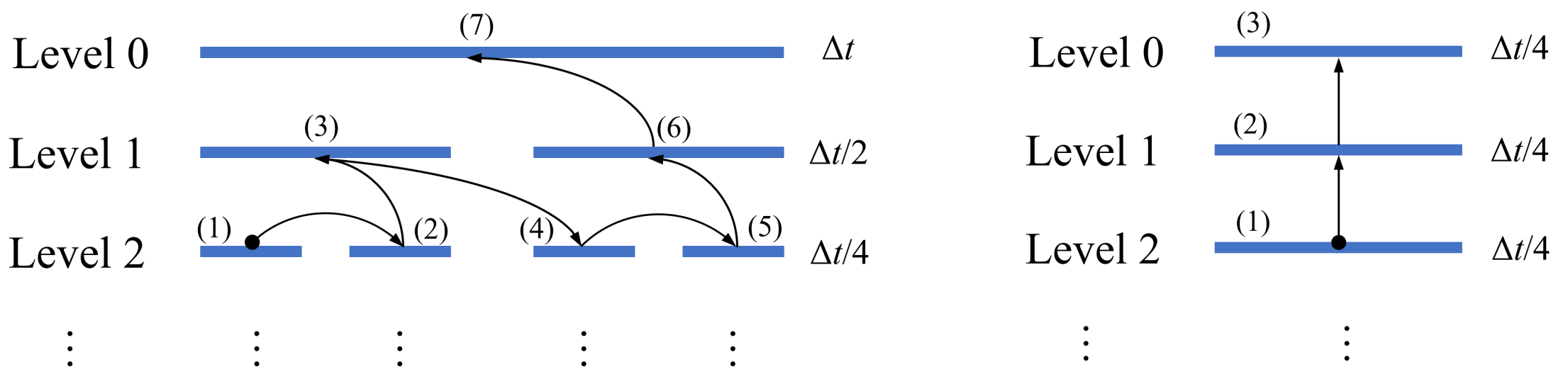

**Figure 2**. Time-advance strategies for the solution within JAX-AMR.

Algorithm 3 further details the implementation algorithm within the `solver` function. The functions invoked in sequence include:

- `get_ghost_block`: This function retrieves the `ghost_blk_data` for `blk_data` based on the neighbor information in `blk_info`.
- `advance_flux`: This function performs flux computations on `ghost_blk_data`, with the specific algorithms related to this process introduced in Section 4.
- `update_physical_boundary`: The external boundaries of blocks in the base layer $B_0$ correspond to the physical boundaries of the computational domain. This function updates the external boundaries of these blocks.
- `update_external_boundary`: For blocks in the refined layers, this function removes ghost grids to restore `blk_data` and synchronizes the external boundary data in `blk_data` with the corresponding data from the previous layer's grids.
- `advance_source_term`: This function performs source term updates on `blk_data`, with the specific algorithms related to this introduced in Section 4.

If a multi-step Runge-Kutta method is employed for time advance, a similar process should be executed at each step.

```
in solver do
    get_ghost_block ()
```

```
    advance_flux ()
    if  level = 0  do
        update_physical_boundary ()
    else  do
        update_external_boundary ()
    end
    advance_source_term ()
end
```

**Algorithm 3**: Implementation details in a solver function combined with AMR

**4. Compressible reacting flow solver**

*4.1 Equations solved by JANC*

The current version of JANC solves the two-dimensional compressible reactive Euler equations in a Cartesian coordinate system $(x, y)$, as expressed in Equation (4.1). The effects of viscosity and transport properties of the working fluid are neglected. In Equation (4.1), the vectors are defined as follows: $\boldsymbol{U} = [\rho, \rho u, \rho v, E, \rho Y_1, \ldots, \rho Y_{Ns\text{-}1}]^T$, $\boldsymbol{F} = [\rho u, \rho u^2 + p, \rho uv, u(E + p), \rho uY_1, \ldots, \rho uY_{Ns\text{-}1}]^T$, $\boldsymbol{G} = [\rho v, \rho uv, \rho v^2 + p, v(E + p), \rho vY_1, \ldots, \rho vY_{Ns\text{-}1}]^T$, $\boldsymbol{S} = [0, 0, 0, 0, \dot{\omega}_1, \ldots, \dot{\omega}_{Ns-1}]^T$. Here, $\rho$, $u$ and $v$ denote the density, $x$-direction velocity, and $y$-direction velocity of the gas mixture, respectively; $E$ is the total energy, $p$ is the pressure, $Y_k$ is the mass fraction of the $k$-th species, $\dot{\omega}_k$ is the production rate of the $k$-th species, and $Ns$ is the total number of species in the mixture.

$$\frac{\partial}{\partial t}\boldsymbol{U} + \frac{\partial}{\partial x}\boldsymbol{F}(\boldsymbol{U}) + \frac{\partial}{\partial y}\boldsymbol{G}(\boldsymbol{U}) = \boldsymbol{S}(\boldsymbol{U}) \tag{4.1}$$

To close Equation (4.1), it is necessary to solve both the thermodynamic equation of state for the gas mixture and the chemical reaction kinetics. JANC employs the thermally perfect gas equation of state, as given in Equation (4.2), where $R$ is the specific gas constant of the mixture, and $h$ is the specific enthalpy of the mixture, calculated as the mass-fraction-weighted sum of the specific enthalpies of individual species $h_k$. In the computation of Equation (4.2), the mixture properties are evaluated using the standard formulation for ideal gas mixtures:

$$\begin{gathered} p = \rho RT, \\ E = \rho h - p + \frac{1}{2}\rho\left(u^2 + v^2\right), \\ h = \sum_{k=1}^{Ns} Y_k h_k \end{gathered} \tag{4.2}$$

In the evaluation of the equation of state, the thermodynamic properties of each gas-phase species (species $k$)—including the specific heat at constant pressure $C_{p,k}$, the specific heat at constant volume $C_{v,k}$, the specific enthalpy $h_k$, and the entropy $s_k$—are computed using the NASA polynomial formulations:

$$\frac{C_{p,k}}{R_u/M_k} = a_{0,k} + a_{1,k}T + a_{2,k}T^2 + a_{3,k}T^3 + a_{4,k}T^4 \quad , \quad C_{v,k} = C_{p,k} - \frac{R_u}{M_k} \tag{4.3}$$

$$\frac{h_k}{R_u/M_k} = a_{0,k}T + \frac{a_{1,k}}{2}T^2 + \frac{a_{2,k}}{3}T^3 + \frac{a_{3,k}}{4}T^4 + \frac{a_{4,k}}{5}T^5 + a_{5,k} \tag{4.4}$$

$$\frac{s_k}{R_u/M_k} = a_{0,k}\ln T + a_{1,k}T + \frac{a_{2,k}}{2}T^2 + \frac{a_{3,k}}{3}T^3 + \frac{a_{4,k}}{4}T^4 + a_{6,k} \tag{4.5}$$

where $R_u$ = 8.314472 J/(mol·K) is the universal gas constant, and $M_k$ (kg/mol) is the molar mass of the $k$-th species. In JANC, the polynomial coefficients $a_{i,\,k}$ are retrieved from the open-source thermodynamic library Cantera [25]. Users also have the flexibility to import custom thermodynamic databases, provided that they are compatible with Cantera.

For chemical reaction kinetics, JANC employs a finite-rate reaction model, capable of simulating elementary reactions based on detailed chemical mechanisms. Specifically, for a reaction system involving $Ns$ species and $Nr$ reactions, the $i$-th elementary reaction can be expressed as:

$$\sum_{j=1}^{Ns} v_{ij}^{f}\chi_j \Leftrightarrow \sum_{j=1}^{Ns} v_{ij}^{b}\chi_j, \quad j = 1,2,\ldots,Nr \tag{4.6}$$

where $\chi_j$ denotes the $j$-th species, and $v_{ij}^{f}$ and $v_{ij}^{b}$ represent the stoichiometric coefficients of the reactants and products, respectively, for species $j$ in the $i$-th reaction. The chemical source term for species $k$ can then be expressed as:

$$\dot{\omega}_k = M_k \sum_{i=1}^{Nr}\left[\left(v_{ik}^{b} - v_{ik}^{f}\right)\left(\sum_{j=1}^{Ns} a_{ij}X_j\right)\left(k_{fi}\prod_{j=1}^{Ns} X_j^{v_{ij}^{f}} - k_{bi}\prod_{j=1}^{Ns} X_j^{v_{ij}^{b}}\right)\right] \tag{4.7}$$

In the above expression, $a_{ij}$ denotes the third-body coefficient of species $j$ in the $i$-th reaction, and $X_k$ represents the molar concentration of species $k$. The forward and backward rate constants, $k_{fi}$ and $k_{bi}$, are defined using the Arrhenius expression in conjunction with the equilibrium constant $K_i^{eq}$ of the reaction, as follows:

$$k_{fi} = A_i T^{b_i} e^{-\frac{E_{a,i}}{R_u T}} \quad , \quad k_{bi} = k_{fi}/K_i^{eq} \tag{4.8}$$

$$K_i^{eq} = \exp\left[\sum_{j=1}^{Ns}\left(v_{ij}^{b} - v_{ij}^{f}\right)\left(\frac{s_j}{R_u/M_j} - \frac{h_j}{R_u T/M_j}\right)\right]\cdot\left(\frac{101325Pa}{R_u T}\right)^{\sum_{j=1}^{Ns}\left(v_{ij}^{b} - v_{ij}^{f}\right)} \tag{4.9}$$

In JANC, Equations (4.1) – (4.9) together form a complete and closed system of equations, enabling the coupled solution of flow dynamics, thermophysical properties based on NASA polynomials, and detailed elementary chemical reactions for a gas mixture defined by user-specified species. JANC includes an input interface capable of reading any reaction mechanism file in chemkin format [26], allowing users to load custom chemical mechanisms for reactive flow simulations.

### *4.2 Numerical methods in JANC*

JANC employs a conservative finite difference method to semi-discretize Equation (4.1), ensuring the conservation properties of the governing equations are preserved during numerical computation:

$$\frac{\partial}{\partial t}\boldsymbol{U}_{i,j} + \frac{\boldsymbol{F}_{i+1/2,j} - \boldsymbol{F}_{i-1/2,j}}{\Delta x} + \frac{\boldsymbol{G}_{i,j+1/2} - \boldsymbol{G}_{i,j-1/2}}{\Delta y} = \boldsymbol{S}_{i,j} \tag{4.10}$$

where the subscripts $i, j$ indicate the location of the solution variables on a Cartesian structured grid, while the 1/2 subscript denotes the interfaces between adjacent grid cells. $\Delta x$ and $\Delta y$ are the uniform grid spacings in the $x$- and $y$-directions, respectively, as JANC uses uniformly spaced grids. JANC adopts the Lax-Friedrichs flux vector splitting (FVS) method, which decomposes the flux into two components propagating in opposite directions in characteristic space. The splitting takes into account the local wave propagation speed, including the local sound speed $c_{i,j}$ at the grid point $(i, j)$:

$$\begin{gathered} \boldsymbol{F}_{i,j} = \boldsymbol{F}_{i,j}^{+} + \boldsymbol{F}_{i,j}^{-}, \\ \boldsymbol{F}_{i,j}^{\pm} = \frac{1}{2}\left(\boldsymbol{F}_{i,j} \pm \alpha \boldsymbol{U}_{i,j}\right), \alpha = \max_{i,j}\left(\left|u_{i,j}\right| + c_{i,j}\right) \end{gathered} \tag{4.11}$$

In reconstructing the flux values at the 1/2 grid interfaces, JANC uses the WENO5 scheme to interpolate and reconstruct the fluxes in the negative and positive normal directions, respectively. The WENO5 method based on flux splitting essentially requires 6 template points (3 on each side) to reconstruct the inflow and outflow fluxes at a grid interface. To avoid the accuracy loss at the physical boundary with the WENO5 scheme, JANC sets up 3 layers of ghost grids at the boundary, so that the flux solver can perform computations with a consistent number of template points within the discretized domain without needing separate local accuracy reduction handling.

For time-advance solutions, if the problem uses a detailed elementary reaction chemical model, in order to eliminate the stiffness of the chemical reaction source terms, JANC employs the Strang Source Splitting method [27] to decouple the convection terms and chemical reaction source terms. The time advancement of the convection terms uses the 3rd-order TVD-Runge-Kutta method, and the chemical reaction source term advancement uses a first-order point-implicit method [28]. The flow field conserved quantities after one time step $\Delta t$ for the convection terms can be expressed as $\boldsymbol{U}_{i,j}^{*}$, and $\boldsymbol{S}_{i,j}^{*} = \boldsymbol{S}\left(\boldsymbol{U}_{i,j}^{*}\right)$ represents the corresponding reaction source terms. Therefore, the first-order point-implicit method can be expressed as:

$$\left(\boldsymbol{I} - \Delta t \frac{\partial \boldsymbol{S}_{i,j}^{*}}{\partial \boldsymbol{U}_{i,j}^{*}}\right) \Delta \boldsymbol{U}_{i,j}^{*} = \boldsymbol{S}_{i,j}^{*} \Delta t \tag{4.12}$$

$$\boldsymbol{U}_{i,j}^{new} = \boldsymbol{U}_{i,j}^{*} + \Delta \boldsymbol{U}_{i,j}^{*} \tag{4.13}$$

In the equation, $\boldsymbol{I}$ represents the identity matrix, $\boldsymbol{U}_{i,j}^{new}$ represents the conserved quantities of the flow field at the next time step, and $\partial \boldsymbol{S}_{i,j}^{*} / \partial \boldsymbol{U}_{i,j}^{*}$ represents the Jacobian matrix of $\boldsymbol{U}_{i,j}^{*}$. In the Jacobian matrix, due to the use of the Source Splitting method, only $\rho Y_1$, …, $\rho Y_{Ns-1}$ are considered as the variables for differentiation, while other conserved quantities are treated as constants.

Additionally, during the solution process, both the flux splitting and the calculation of the chemical reaction source terms require the temperature to be inferred from the enthalpy of the flow field. JANC uses the Newton-Raphson iteration method to solve the quartic polynomial equation (4.4) for enthalpy and temperature. To accelerate the convergence of the iteration, the initial temperature for each Newton-Raphson iteration is set as the temperature from flow field of the previous time step. This way, the iteration algorithm only needs to fine-tune the temperature change within a single time step.

### *4.3 Implementation details in JANC*

This section introduces the implementation of the numerical algorithms from Section 4.2 in JANC

from the perspective of code organization and algorithm design. The main body of the JANC program follows a modular programming approach, storing functions with different functionalities in separate **.py** submodules. An overview of the program modules is shown in Table 2.

**Table 2** File structure of current version of JANC

JANC/src

| | |
|---|---|
| load.py | # load thermo and chemical mechanism **input from user** |
| thermo.py | # thermodynamics properties calculation (gamma, R *et al.*) |
| chemical.py | # reaction rate and point implicit Jacobian calculation |
| flux.py | # Lax-Friedrich flux splitting and WENO5 reconstruction |
| boundary.py | # **user-defined** boundary conditions |
| aux_func.py | # **user-defined** source terms and other auxiliary functions |
| nondim.py | # **user-defined** non-dimensionalization parameters |
| cfd_solver.py | # main solver which calls the above modules |
| parallel_solver.py | # parallel implementations of the main solver across GPUs |

On the right side of Table 2, an introduction to the functions of each module is provided, and JANC offers 4 user-customizable interfaces that can be modified according to the user's problem. These interfaces are highlighted in bold in the function descriptions:

1. In **load.py**, users can load custom thermodynamic databases and detailed chemical reaction mechanisms, as long as the data format is consistent with Cantera's requirements.
2. In **boundary.py**, users can customize boundary condition functions through 4 functions: `left_boundary`, `right_boundary`, `bottom_boundary` and `up_boundary`. The content of this function is entirely user-defined, as long as it accepts the flow field state variables and outputs the flow field state variables with 3 layers of ghost grids on both the *x*- and *y*- sides.
3. In **aux_func.py**, users can add custom equation source terms in the `source_terms` function. These source terms will be advanced along with the convection terms using the TVD Runge-Kutta 3rd-order time-advance algorithm. For example, if the user's problem does not require detailed chemical reaction mechanisms but involves a weaker stiff global reaction, the source terms for this chemical reaction can be directly added to the `source_terms` function and advanced together with the convection terms, without the need to use the point-implicit method for splitted time-advance. When users write their custom functions, they can freely call functions from modules such as **thermo.py** to assist in implementing their functionality.
4. In **nondim.py**, users can customize the nondimensionalization constants of the equations. In JANC, the default solving process is nondimensionalized, with the aim of making the physical quantities solved by the solver have comparable magnitudes, facilitating the post-processing of the computed data. For example, when combining with downstream machine learning tasks, there is no need for additional normalization of the solver data.

The functions of each submodule are integrated in **cfd_solver.py**, which implements the spatial discretization and time-advance as described in Section 4.2. Therefore, this section will mainly focus on the implementation of **cfd_solver.py**, while other modules' functions and their functionalities can be reviewed in the documentation, source code, and comments on GitHub.

**cfd_solver.py** mainly introduces the following core functions from other modules:

- `boundary.boundary_conditions`: This function integrates the user's 4 custom functions (which define the boundary conditions for the left, right, top, and bottom sides). It accepts the flow field conserved quantity `U` and thermodynamic state variables `aux` ($\gamma$ and $T$) without ghost cells, and outputs `U_with_ghost_cells` and `aux_with_ghost_cells` that include the ghost cells.
- `flux.weno5`: This function receives `U_with_ghost_cells` and `aux_with_ghost_cells`, and through LF splitting and WENO5 reconstruction, outputs the flux gradient `-(dF/dx + dG/dy)` without the ghost cells.
- `aux_func.source_terms`: This function is a user-defined source term function (if no custom source term is defined, the output of this function is 0). It receives `U` and `aux`, and outputs a source term with the same shape as `U`.
- `aux_func.update_aux`: This function uses the enthalpy values of the current flow field and selects the temperature stored in the thermodynamic quantities `aux` from the previous time step as the initial value for the Newton-Raphson iteration. It then calculates the new temperature `T_new` corresponding to the updated enthalpy and the corresponding specific heat ratio `gamma_new` at that temperature. The function outputs `aux_new = (gamma_new, T_new)`.
- `chemical.solve_implicit_rate`: This function, based on the temperature `T`, molar concentration `X`, and the specified time step `dt`, outputs the calculation of $\Delta U^{*}_{i,j}$ `(dU)` using the implicit point method for equation (4.12).

The **cfd_solver.py** defines the following functions that integrate the functionalities of other modules:

- `cfl`: This function receives `U` and `aux`, and based on the user-specified CFL number, it calculates the time step `dt` that satisfies the CFL condition, using the maximum characteristic speed in the solution domain and the grid size.
- `rhs`: This function receives `U` and `aux`, calls `aux_func.update_aux` firstly to solve the current thermodynamic states from the previous states `aux.` And then it calls `boundary.boundary_conditions` to fill the ghost cells at the physical boundaries, and its output is received by `flux.weno5` to compute the flux gradient. The flux gradient and `aux_func.source_terms` are then summed and output as the right-hand side of the semi-discrete equation. If the solver's computation is performed on an AMR grid's refinement layer, it calls the `get_ghost_block` function from the AMR module introduced in Section 2 to fill the boundaries of the solution domain.
- `advance_flux`: This function advances `U` using the TVD-3rd-order Runge-Kutta method. During each intermediate step, it directly calls the `rhs` function and the `update_aux` function. The output of this function is the result after advancing `U` by one time step `dt`.
- `advance_source_term`: This function first updates the old `aux` with `aux_func.update_aux`, and then the solved `T` and `X` are passed to `chemical.solve_implicit_rate` to get chemical source term `dU` using point-implicit method. It returns `U + dU` to advance one chemical time step.

By calling the above functions, the `advance_one_step` function in **cfd_solver.py** completes the advancement of one time step:

```python
def advance_one_step():
    CFL()
    advance_flux()
    advance_source_term()
```

According to the user's settings, if the chemical reaction source terms are already included in the custom source term function, the last step operation, `advance_source_term`, will not be executed. In this case, the solver will call the explicit version of `advance_one_step`:

```python
def advance_one_step():
    CFL()
    advance_flux()
```

To further improve the computational speed of JANC, we have also developed parallel computation functionality. For multi-GPU systems, JANC with parallel functionality enabled can evenly distribute the grid of the computational domain across different GPUs. The parallel strategy is shown in Figure 3. In the figure, N GPUs evenly distribute the computational domain (including physical boundary conditions) along the $x$-axis, and each GPU only needs to handle 1/N of the total grid. The left and right boundary conditions are handled only by GPU_0 and GPU_N-1, respectively. During each time step, the GPUs send boundary grids to neighboring GPUs for communication of internal boundary grids. Currently, JANC's parallel partitioning strategy is relatively simple, with grid allocation only along one direction. More complex and efficient parallel partitioning strategies will be developed in the future.

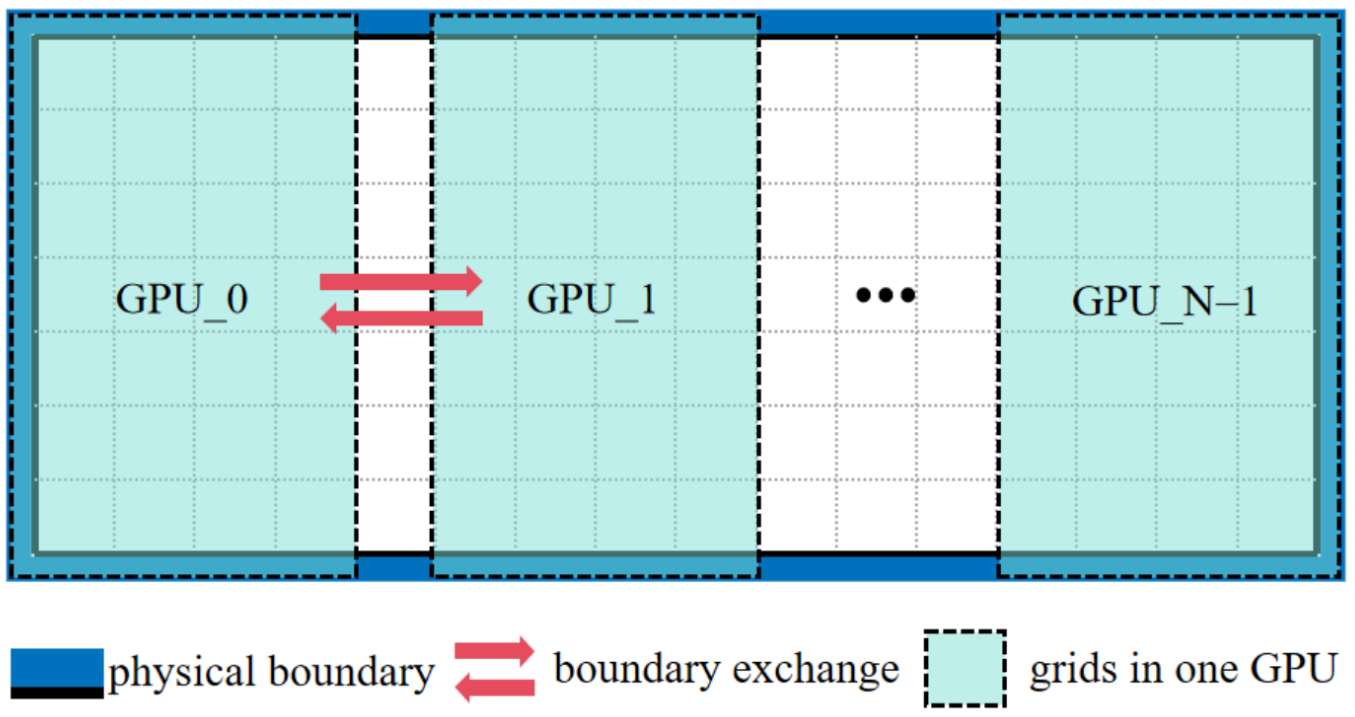

**Figure 3**. Grid Partitioning Strategy for Parallel Computing in JANC

The parallel functionality of JANC is integrated in the **parallel_solver.py** file, with the following core functions:

- `split_grid`: This function divides the computational domain's grid into multiple sets of grids along the $x$-axis based on the number of GPUs, and assigns different grids to different GPUs for processing.
- `exchange_halo`: This function sends the left and right boundary data of the grids being processed by the current GPU to the left and right neighboring GPUs, respectively, while receiving the boundary data from the two neighboring GPUs. This facilitates communication between neighboring grid points on different GPUs. The communication functionality is implemented using the `jax.lax.ppermute` function.
- `parallel_boundary_conditions`: This function imports the user-defined physical boundary conditions for single-GPU computations, assigns the left and right boundary conditions to GPU_0 and GPU_N-1, and assigns the top and bottom boundary conditions to all GPUs.

- `parallel_rhs`: Similar to `rhs` in cfd_solver.py, but with the single-GPU `boundary_conditions` replaced by `parallel_boundary_conditions`.
- `parallel_advance_one_step`: Similar to `advance_one_step` in **cfd_solver.py**, but with `rhs` replaced by `parallel_rhs`, and the `@jax.pmap` decorator added to the function declaration.

The convenient implementation of these parallel functionalities is another advantage of the JAX framework solver. In **parallel_cfd_solver.py**, apart from the application of boundary conditions and communication parts, other functions directly call the single-GPU functions from **cfd_solver.py**. By simply adding the `@jax.pmap` decorator, the program can automatically implement parallelism without the need to separately write MPI files, as is typical in traditional CFD software. Even when parallel-specific boundary conditions need to be adjusted, JANC's parallel boundary conditions directly call the single-GPU boundary condition functions, so the user-defined functions do not need to be modified for parallelism.

## 5. Validation and Performance analysis of JANC

This section demonstrates the capability of JANC in solving several typical compressible reacting flow and combustion problems, and compares its results with those from existing main-stream open-source solvers for validation and analysis. In Section 5.1, the accuracy of JANC's core solver is verified using benchmark cases. Section 5.2 presents a performance comparison between JANC and OpenFOAM, highlighting the computational efficiency of the core solver, as well as the parallel efficiency and the improved performance when integrated with JAX-AMR.

### *5.1 Validations of the solver*

#### (1) SOD Shock tube tests

This section presents a verification test using a one-dimensional shock tube problem, comparing the solver for convective terms of JANC with the analytical solution of the Riemann problem. The computational domain is set to $x \in [-10, 10]$, with non-reflecting boundary conditions on both sides. The thermodynamic model adopts an ideal gas with a constant specific heat ratio, $\gamma = 1.4$, and gas constant $R = 287$ J/(kg • K).The initial conditions for the Riemann problem are: For $x < 0$: $\rho = 1$ kg/m$^3$, $u = 0$ m/s, $p = 10^5$ Pa; For $x \geq 0$: $\rho = 0.125$ kg/m$^3$, $u = 0$ m/s, $p = 10^4$ Pa.

Since this is a one-dimensional validation case, only a single layer of grid is used in the $y$-direction. The $x$-direction is discretized with 400 uniformly spaced cells. Non-reflecting boundary conditions are applied in both directions. Chemical reactions are not included in this test; only the conservation equations of mass, momentum, and energy are solved. The CFL number for time integration is set to 0.2.

At $t = 0.01$, the flow field variables—density, velocity in the $x$-direction, and pressure—are extracted and compared against the analytical solution of the Riemann problem. As shown in Figure 4, the results obtained by JANC's solver closely match the analytical solution. The solver accurately captures discontinuities and expansion waves with high resolution on the selected grid.

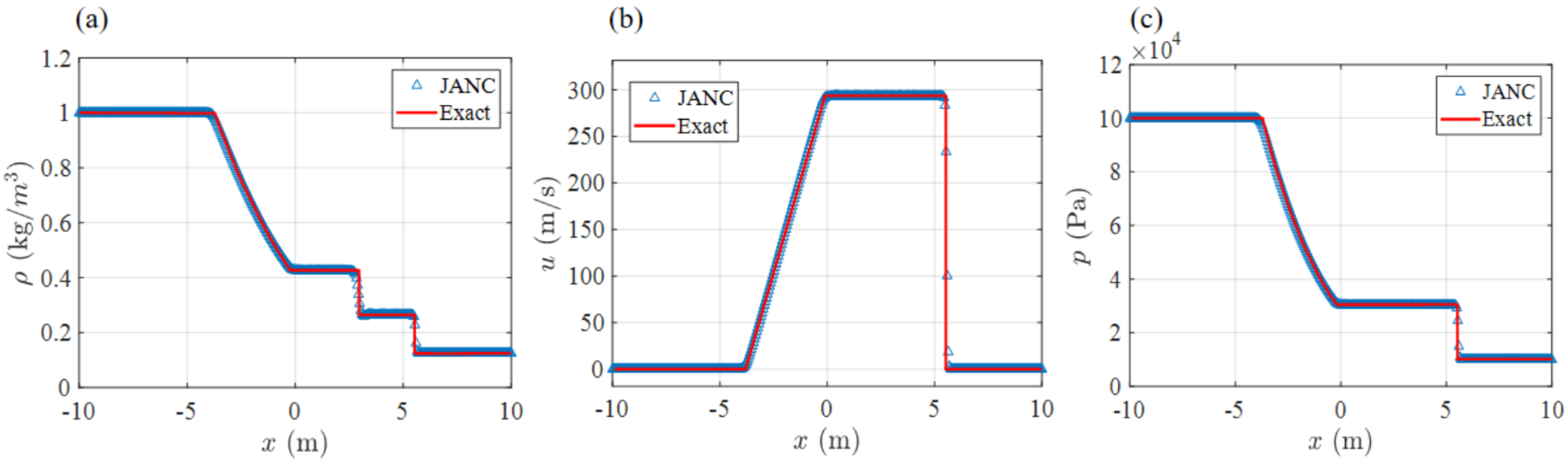

**Figure 4**. Profiles of the shock tube problem at $t$ = 0.01s: (a)density; (b)velocity; (c)pressure.

**(2) Ignition delay tests**

In addition to verifying the accuracy of the convective solver, it is also essential to assess the correctness of the chemical reaction kinetics, including the stability and accuracy of the point-implicit algorithm. In this section, we select the ignition delay time under specified initial pressure and temperature in a zero-dimensional constant-volume reactor as the validation metric. The results computed by JANC are compared with those from Cantera using the same reaction mechanism. We employ Jachimowski's detailed hydrogen-oxygen mechanism, consisting of 9 species and 19 reactions [29]. The mechanism is written in the chemkin format with the following unit system: cm-s-mol-cal-K. In JANC, the **.inp** mechanism file is directly read, and the thermodynamic polynomial data are taken from NASA-7 coefficients in `gri30.yaml`. For Cantera, the relevant 9 species' thermodynamic coefficients are extracted from `gri30.yaml` and written into a `thermo.dat` file in chemkin format. Together with the **.inp** file, this data is used as input to Cantera's `ck2yaml` utility, generating a **.yaml** file that can be loaded by the `Solution` class in Cantera.

To simulate a zero-dimensional constant-volume reactor, both the $x$ and $y$ boundaries in JANC are set to non-reflecting, with only one cell in each direction to freeze the convective terms. In Cantera, `IdealGasReactor` is used to model the same physical condition. Both JANC and Cantera adopt identical initial pressure, initial temperature, and simulation end time. JANC uses a point-implicit method with fixed time step size, whereas Cantera utilizes its default solver with adaptive time stepping.

Two validation cases are used for comparison in this section: **Case 1**: Initial pressure $P$ = 20 atm, initial temperature $T$ = 1500 K. The temperature evolution over time computed by both solvers is compared; **Case 2**: Initial pressure fixed at $P$ = 20 atm, and ignition temperature varied across 9 values $T$ = [1200 K, 1300 K, …, 2000 K]. The ignition delay times predicted by both solvers are compared.

To select an appropriate time step `dt` and verify the stability of the point-implicit scheme, Figure 5 compares the temperature-time profiles of **Case 1** under three different time steps. The results show that time steps on the order of $10^{-9}$ and $10^{-8}$ provide stable results for elementary reactions. However, a larger time step of $5\times10^{-8}$ leads to noticeable deviation from the other two time steps. Therefore, a time step of `dt = 5e-9` is adopted for the following tests.

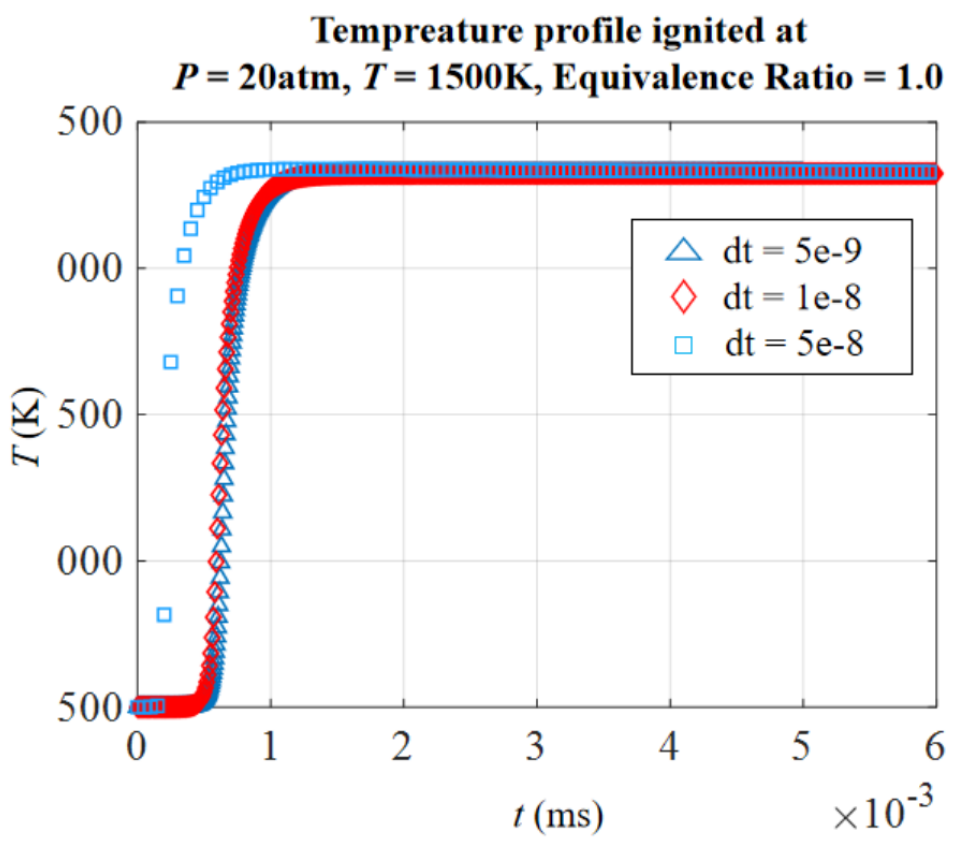

**Figure 5**. Simulation results of **Case 1** under different time step sizes.

The comparisons between JANC and Cantera for the two validation cases are shown in Figure 6. Both the temperature evolution after ignition and the ignition delay times predicted by JANC are in good agreement with those computed by Cantera. This indicates that the chemical reaction kinetics module in JANC—including both the reaction rate calculations and the point-implicit integration method—is accurate and reliable.

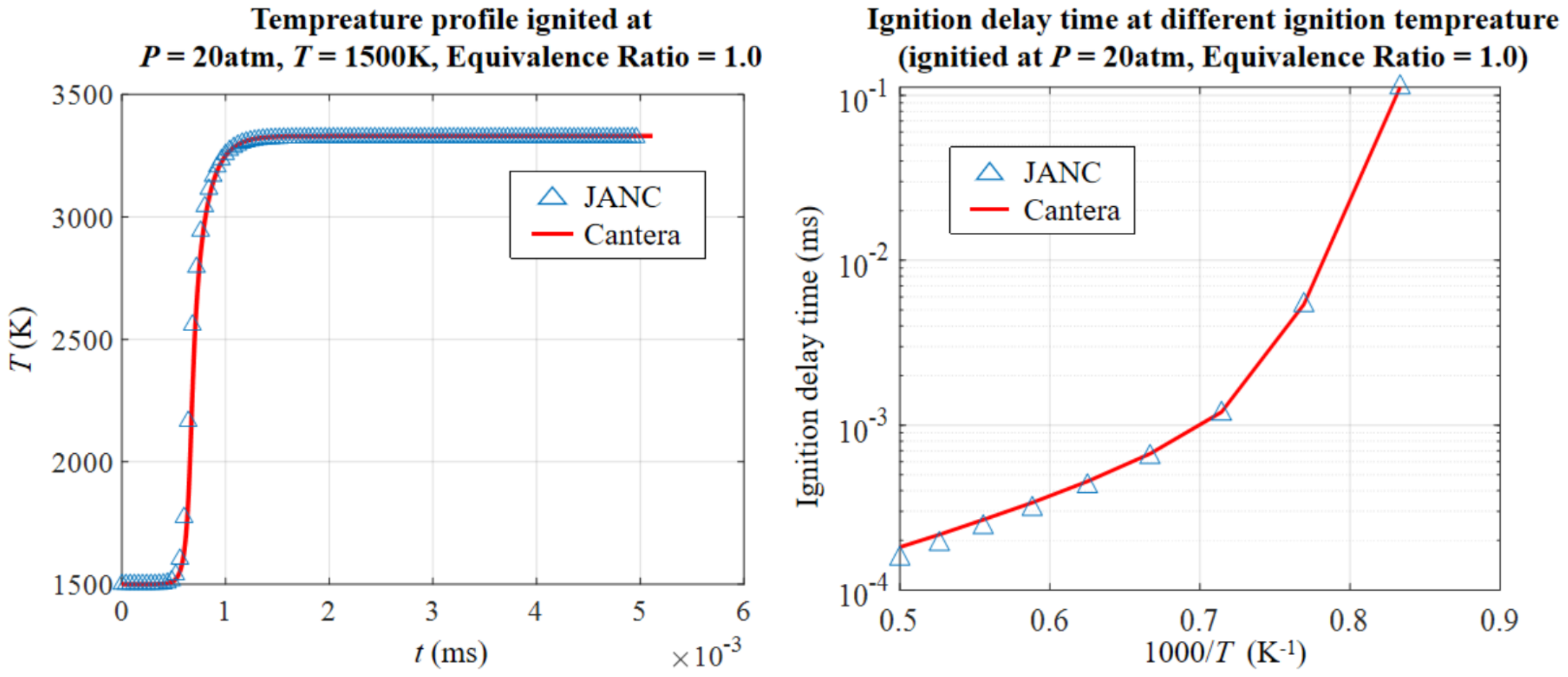

**Figure 6**. Results of **Case 1** (left) and **Case 2** (right).

*5.2 Performance Analysis*

This section uses a two-dimensional detonation tube case to evaluate the computational performance of JANC's core solver, parallelization capabilities, and AMR functionality. The results are compared and analyzed against those obtained from OpenFOAM (specifically, detonationFOAM solver [30] is used).

*5.2.1 Set-up of detonation tube simulation*

As shown in Figure 7, the two-dimensional detonation tube in the test case has dimensions of $L_x$ = 0.05 m and $L_y$ = 0.0125 m. A premixed region and an ignition region are defined within the tube. The premixed region is filled with a uniformly mixed $H_2$-$O_2$-$N_2$ gas mixture, with mass fractions of $H_2 : O_2 : N_2 = 0.028 : 0.226 : 0.746$, at an initial temperature $T_0 = 300$ K and pressure $P_0 = 1$ atm. On the left end of the detonation tube, the ignition zone has a width of $w_i = 0.002$ m. To facilitate detonation initiation, three semi-circular ignition zones are evenly distributed along the $y$-direction, each with a radius of $r_i = 0.001$ m and spacing $h_i = 0.003125$ m. The composition within the ignition zones is the same as in the premixed region, but with an elevated temperature $T_i = 3500$ K and pressure $P_i = 75$ atm. All four boundaries of the computational domain are set as slip wall boundaries. The 9-species 19-step detailed hydrogen-oxygen mechanism [29], the same as in Section 5.1, is used for combustion chemistry.

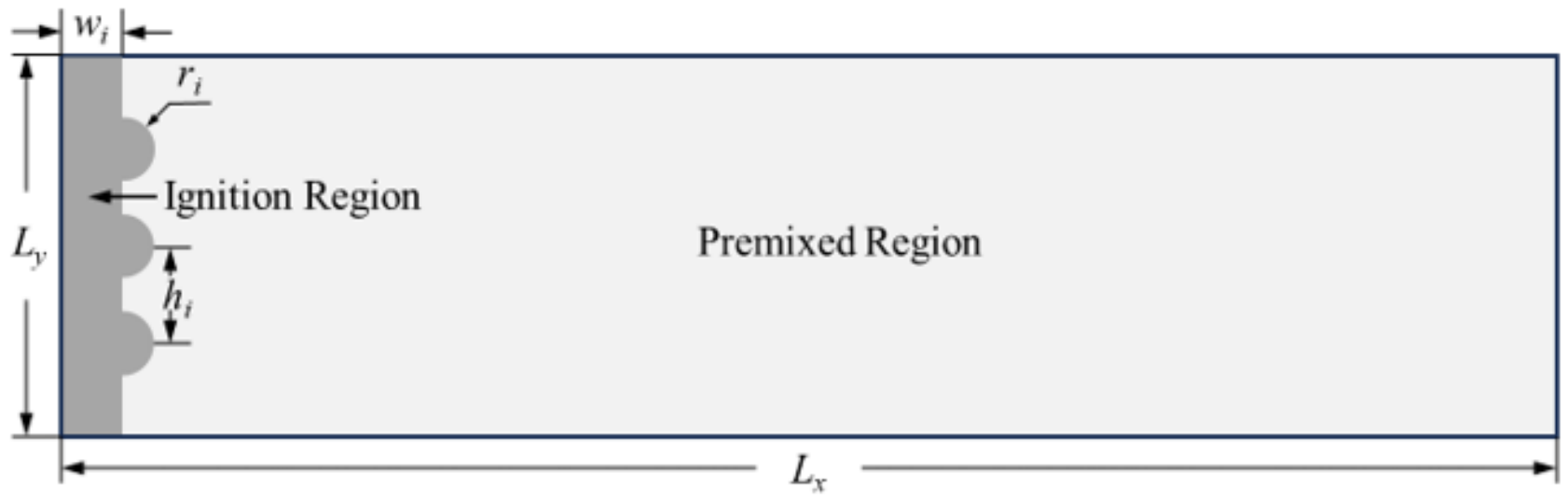

**Figure 7**. Schematic diagram of the computational domain for the detonation tube test case.

For the JANC simulation, the WENO5 scheme is used for spatial discretization, with third-order Runge-Kutta time stepping. The chemical source terms are handled with a first-order point-implicit method, as described in detail in Section 4. Correspondingly, in the OpenFOAM simulation, the second-order upwind MUSCL scheme with a minmod limiter is used, combined with an HLLC-P approximate Riemann solver (HLLC with a pressure-control technique). An operator-splitting method decouples the flow terms and chemical source terms. The flow solver uses a first-order implicit Euler scheme for time integration, while the chemical reactions are integrated using the seulex ODE solver.

JANC simulations are run on an NVIDIA A100 GPU with a double-precision (FP64) performance of 9.7 TFLOPS. OpenFOAM simulations are executed using multi-core AMD EPYC 7452 CPUs running at 2.35 GHz.

It should be noted that, since the cell size is closely related to the chosen chemical mechanism, this section only compares the results and computational efficiency of JANC and OpenFOAM under similar configurations. The consistency of the results with experimental data is not discussed here.

*5.2.2 Performance of JANC core solver*

This section analyzes the single-GPU and parallel computational performance of JANC's core solver. The test case uses a uniform mesh with $N_x = 4000$ and $N_y = 1000$ grids in the $x$- and $y$-directions, respectively, resulting in a grid size of 12.5 μm. The simulation time step is set to $\Delta t = 0.5\times10^{-9}$ s。

Figure 8 shows the comparison results at $t = 10$ μs using double-precision computations. Since JANC and OpenFOAM adopt different numerical methods, the detonation front structures obtained from the two solvers are not exactly the same. However, it can be observed that JANC demonstrates superior shock-capturing ability and clearer resolution of the induction zone compared to OpenFOAM.

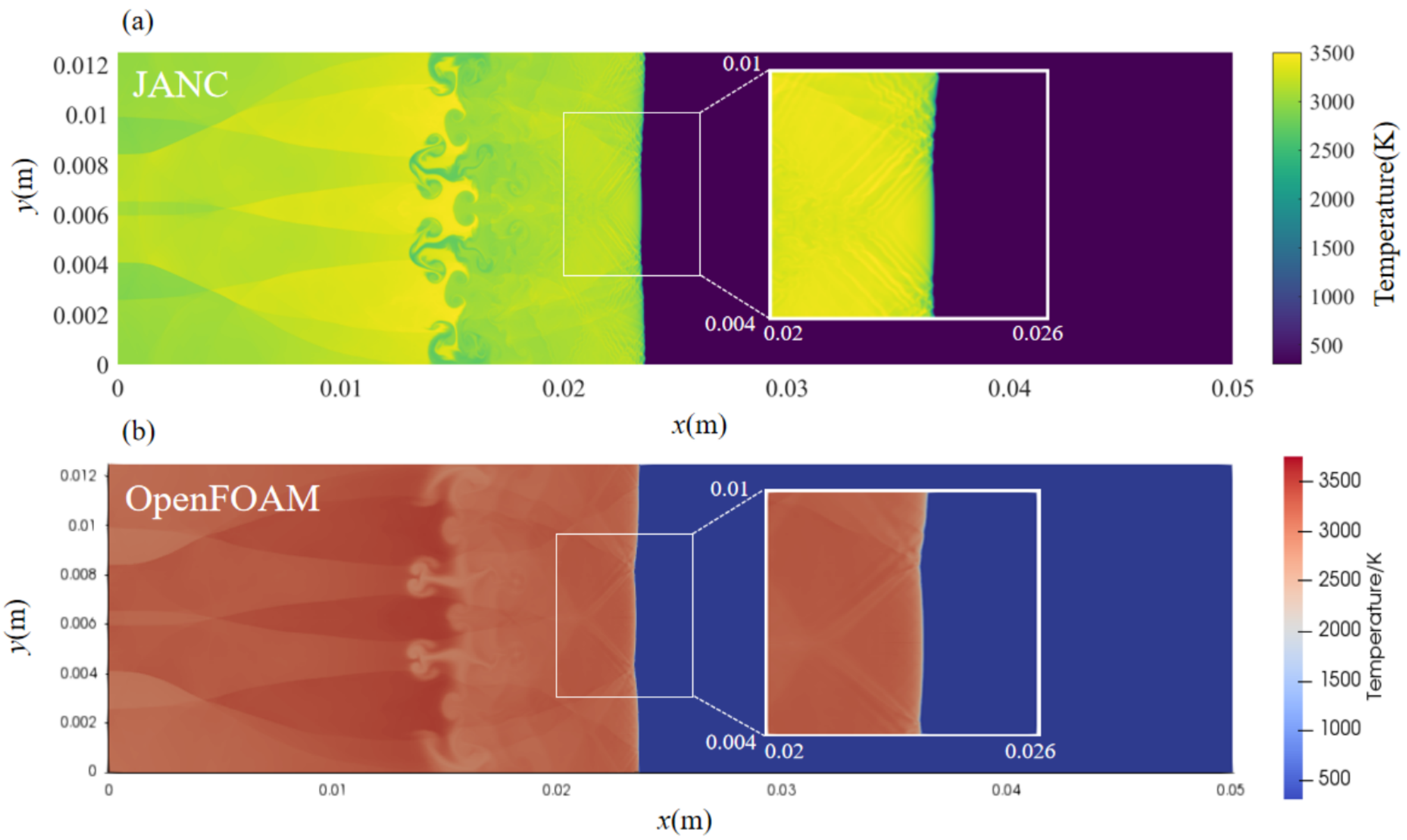

**Figure 8**. Temperature contour plot of the double-precision calculation results: (a) JANC with one A100 GPU; (b) OpenFOAM with 384 AMD CPUs.

In terms of computational efficiency, the JANC case was run on a single A100 GPU, while the OpenFOAM case was run using 384-core AMD CPUs. The average computational speed for JANC is 5.0 step/s, while OpenFOAM achieves 0.5 step/s. Considering current rental prices—approximately 0.24 USD per A100 GPU per hour and 0.007 USD per EPYC 7452 CPU core per hour—the cost per million simulation steps for JANC and OpenFOAM under this test case is respectively

$$\mathrm{CPMS_{JANC}} = 13.3\ \mathrm{USD},\ \mathrm{CPMS_{OpenFOAM}} = 1344.0\ \mathrm{USD}$$

This indicates that the computational cost of JANC is only 1.0% of that of OpenFOAM.

Figure 9 presents the comparison results using 4 GPUs (Figure 9a) and the average computation speeds when using 1, 2, 3, and 4-card A100 GPUs (Figure 9b). It shows that the parallel computing results of JANC are identical to those obtained with a single GPU, and the average computation speed increases almost linearly with the number of GPUs. This indicates that JANC's parallelization introduces negligible changes to the cost per million steps (CPMS), allowing users to significantly shorten computation time by leveraging more GPUs without increasing the overall cost.

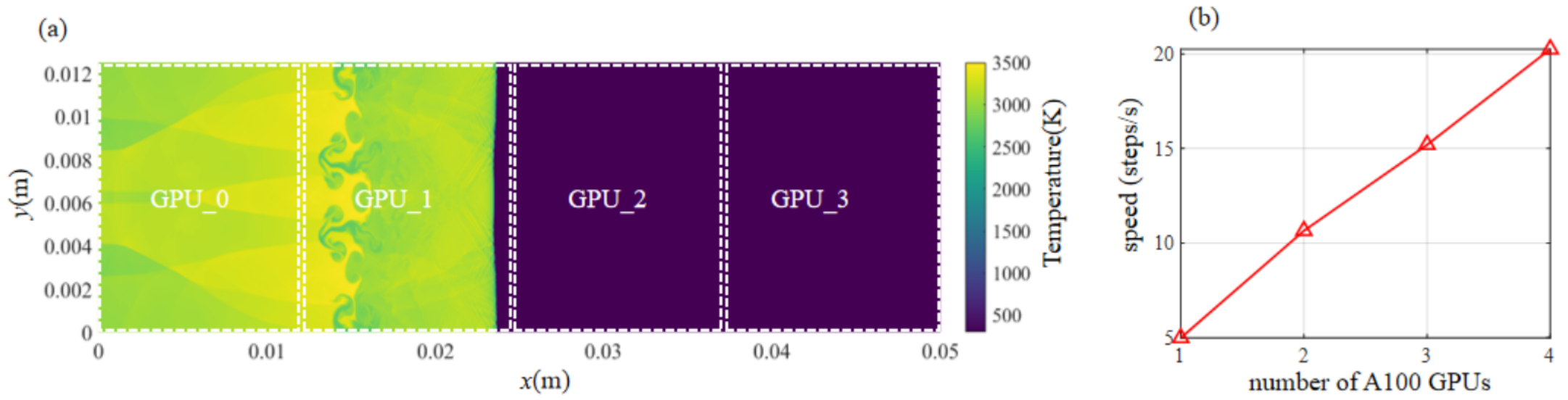

**Figure 9**. (a) Grid distribution and computation results with four A100 GPUs ($t$ = 10 μs); (b) Variation of average computational speed with the number of GPU devices.

### *5.2.3 Performance of solver conjunction with AMR*

The computational performance of the JANC solver in conjunction with JAX-AMR is analyzed in this section. In the comparative example, OpenFOAM solver employs its built-in AMR method based on octree-like data structure [17]. In the computation, the base layer has a grid size of $N_x$ = 2000, $N_y$ = 500, with a grid size of 25 μm. Three refinement layers are used, with the third layer having a gird size of 3.125 μm. To maintain consistency with OpenFOAM, the JANC case adopts the synchronous advance strategy as shown in Figure 2b, with a uniform time step of $\Delta t = 0.125\times10^{-9}$ s for each layer. The AMR update step is set to 40 steps, and the refinement criterion is based on the density gradient criterion. The JANC case utilizes single A100 GPU, while the OpenFOAM case is computed using 384-core AMD CPUs. The computational precision is double precision.

Figure 10 illustrates the results at $t$ = 10 μs. Figure 10a displays the levels of mesh refinement in both the JANC and OpenFOAM cases. To better compare computational performance, the refinement thresholds in both cases are adjusted to ensure their refined areas and total number of refined grids are similar. Figure 10b presents temperature contour plots of the computational results along with a local magnification. It can be observed that due to the use of higher-order discrete schemes and time advance scheme, JANC outperforms OpenFOAM in the detailed identification of the detonation wavefront structure.

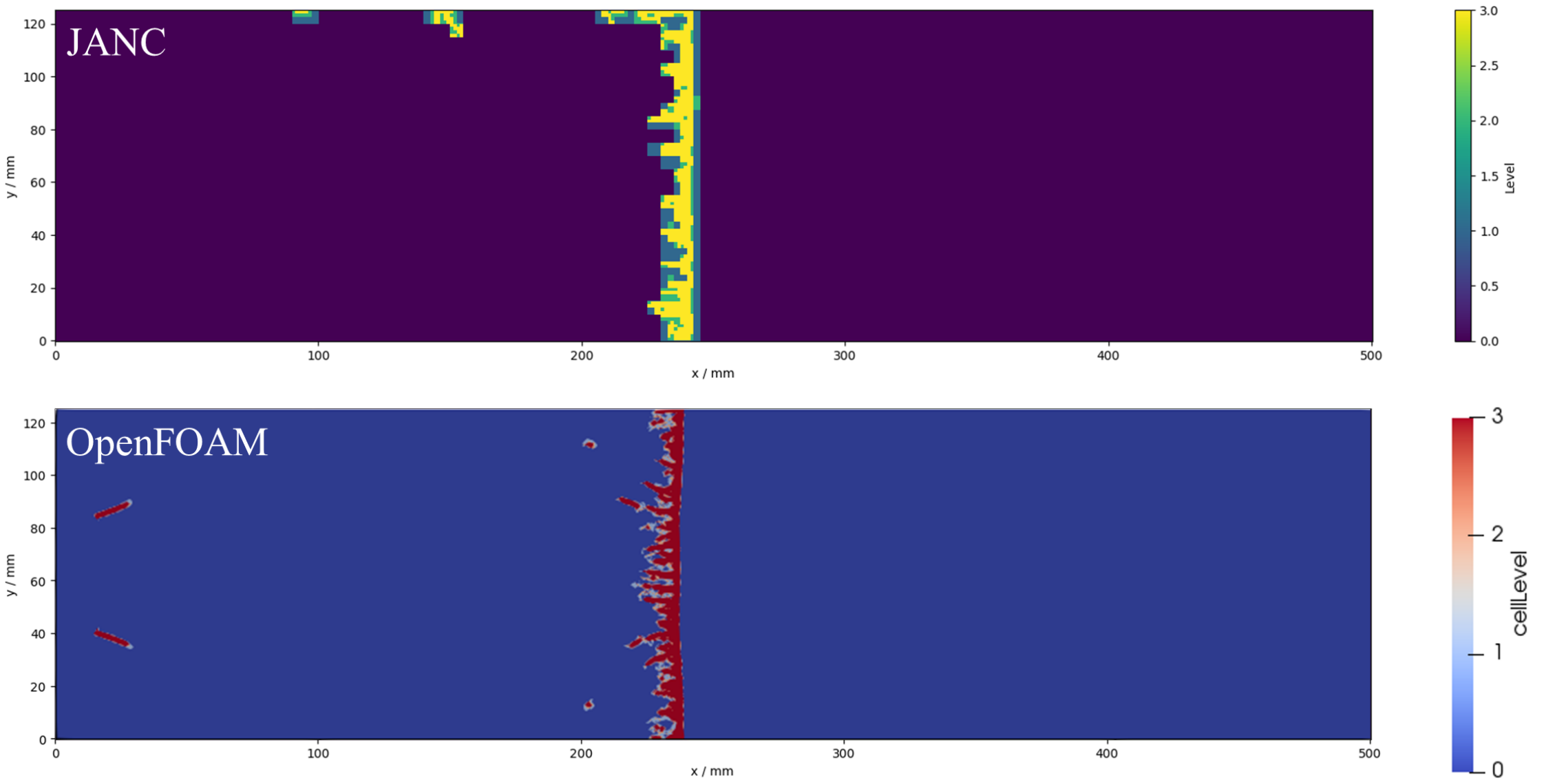

(a) Refinement level for grids and blocks

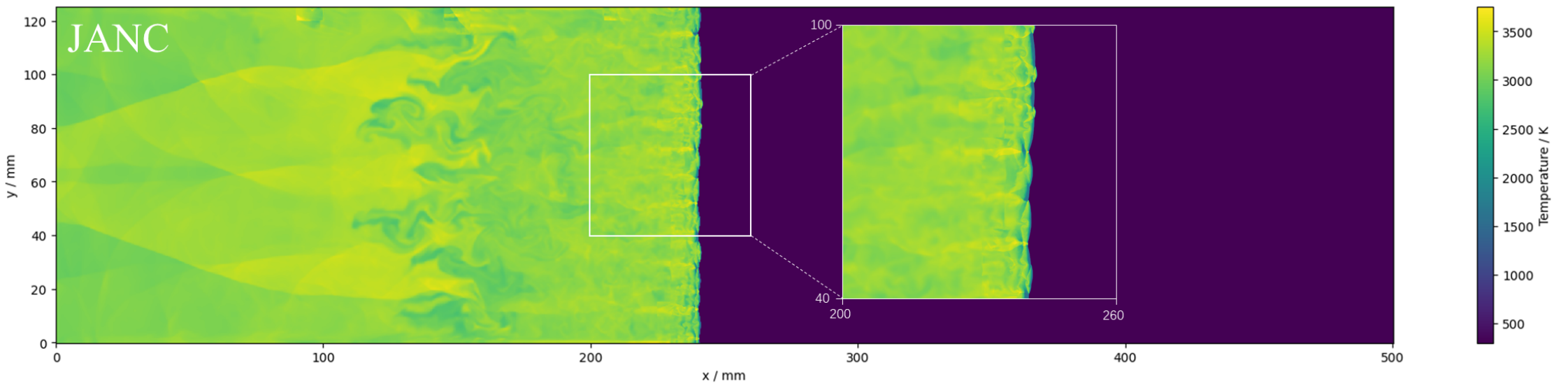

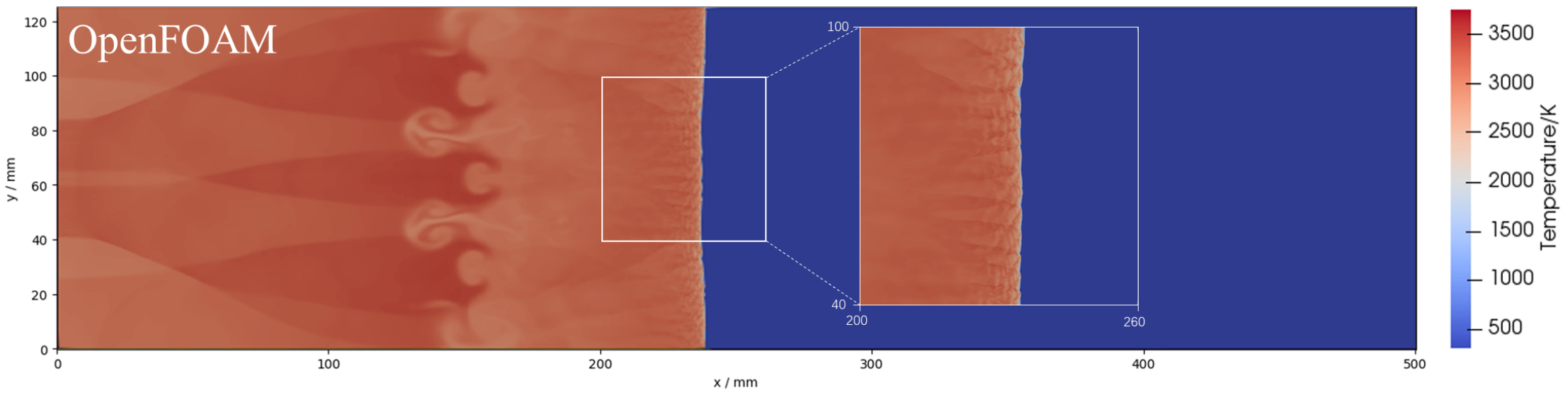

(b) Temperature contour

**Figure 10.** Simulation results ($t$ = 10 μs) using AMR by JANC with single A100 and OpenFOAM with 384-core CPUs: (a) Refinement level; (b) Temperature contour.

In terms of computational efficiency, the total number of grids in all blocks in the JANC case is 2.76 million (including invalid blocks), while the total number of grids in the OpenFOAM case is 2.23 million. The average computation speed for JANC is 0.18 s/step, compared to 0.75 s/step for OpenFOAM. In this comparative case, the CPMS for JANC and OpenFOAM are as follows:

$$\text{CPMS}_{\text{JANC}} = 12.0 \text{ USD}, \text{CPMS}_{\text{OpenFOAM}} = 555.8 \text{ USD}$$

This indicates that JANC's computational cost is only 2.1% of that of OpenFOAM.

It can be observed that compared to the JANC core solver without AMR, there is a slight increase in computational cost when AMR is enabled. This is mainly due to two reasons. First, JAX-AMR employs a block-structured data format that may result in some redundant invalid blocks, and even in valid blocks, there may be some grids that do not need to be refined. Consequently, for the same refinement criterion, the total number of refined grids in JANC will be higher than that of the OpenFOAM case. Second, JANC recommends employing the crossover advance strategy indicated in Figure 2a. Using a synchronous advance strategy introduces additional computational overhead for unnecessary data interpolation between coarse and fine layers. In this example, when JANC adopts the crossover advance strategy, the computational time for advancing one large step ($8\Delta t$) is 0.83 s/step, while the average time for advancing $\Delta t$ is 0.10 s/step. Compared to the synchronous advance strategy, this reduces the computational time by an additional 44%, bringing the computational cost down to 1.2% of that of OpenFOAM.

Furthermore, it should be noted that the computational load of JANC is directly related to the number of blocks, while the dynamic update strategy for $N_{i,\max}$ currently employed in Algorithm 1 of JAX-AMR is not yet fully optimized. This may result in some unnecessary computation and compilation overhead, leading to slower computation speeds. Users can customize the update strategy based on practical requirements, and we will release the optimized update strategy as soon as possible.

## 6. Differentiable solver and adjoint optimization

### *6.1 Implementations of adjoint optimization in JANC*

All functions used in JANC are implemented using the JAX framework, which inherits JAX's automatic differentiation capability. As a result, the solver can be directly embedded into gradient-based end-to-end optimization tasks, including those involving machine learning. The flux solver functions, source term functions, and others in the solver can be directly called by algorithms such as machine learning, participating in tasks like the calculation of the loss function for Physics-Informed Neural Networks (PINN) [31], data-driven numerical discretization learning [32], and flow field reconstruction [13], as well as solving other inverse problem in data-driven approaches.

However, in combustion flow problems, transient dynamic processes are often a key area of research. This means that the solver's time-advance integration loop also needs to be incorporated into the gradient optimization process, allowing the dynamic trajectory of the combustion flow to be traced by the gradient values in the gradient descent algorithm. For convenience, we represent the combustion flow equations in the following general form:

$$\frac{\partial \boldsymbol{U}}{\partial t} = \boldsymbol{N}\left(\boldsymbol{U};\boldsymbol{\theta}\right) \tag{6.1}$$

where $\boldsymbol{U}$ represents the state variables in the combustion flow, and $\boldsymbol{N}$ can either be the Euler equations or a neural network that encompasses the JANC solver. $\boldsymbol{\theta}$ refers to the parameters of the equations, such as the parameters to be optimized in the neural network, boundary condition parameters, thermodynamic property parameters, and so on. The equation (6.1) can be integrated over time to obtain the trajectory of $\boldsymbol{U}$ as a function of time $t$ and parameters $\theta$, i.e., $\boldsymbol{U}(t;\boldsymbol{\theta})$:

$$\boldsymbol{U}\left(t;\boldsymbol{\theta}\right) = \int_{t_0}^{t} \boldsymbol{N}\left(\boldsymbol{U}\left(\tau\right);\boldsymbol{\theta}\right) d\tau + \boldsymbol{U}(t_0;\boldsymbol{\theta}) \tag{6.2}$$

In optimization problems involving differentiable solvers, there is often an optimization cost function $\mathcal{L}$ that depends on $\boldsymbol{U}(t;\boldsymbol{\theta})$. The objective of the optimization problem is to find the parameter $\boldsymbol{\theta}^*$ that minimizes $\mathcal{L}$:

$$\boldsymbol{\theta}^* = \arg\min_{\theta} \mathcal{L}\left(\boldsymbol{U}\left(t;\boldsymbol{\theta}\right)\right) \tag{6.3}$$

To solve for $\boldsymbol{\theta}^*$, a common approach is to use gradient descent optimizers, such as the Adams optimizer, which iteratively optimizes the Jacobian of the cost function $\mathcal{L}$ with respect to the optimization parameters $\boldsymbol{\theta}$:

$$\boldsymbol{\theta}^{n+1} = \boldsymbol{\theta}^{n} + \Delta\boldsymbol{\theta}, \Delta\boldsymbol{\theta} = \text{optimizer}\left(\mathcal{L}, \frac{\partial \mathcal{L}}{\partial \boldsymbol{\theta}}\right) \tag{6.4}$$

To solve for $\partial\mathcal{L}/\partial\boldsymbol{\theta}$, the differentiability of equations (6.1) and (6.2) must be satisfied. For (6.1), the differentiability requirement demands that the flux discretization function and the chemical reaction source term function of the solver be differentiable. This can be achieved by directly inheriting the automatic differentiation properties of JAX functions.

For (6.2), the differentiability requires performing differential calculations on the TVD-RK3 numerical integration process with many time steps. However, directly using JAX's automatic differentiation functions would result in unmanageable memory usage during gradient calculations. This is because the value of the cost function $\mathcal{L}$, determined by $\boldsymbol{U}(t;\boldsymbol{\theta})$, is obtained by iterating over $N_t$ time steps. The state at these intermediate time steps is also dependent on $\boldsymbol{\theta}$, leading to the involvement of all these intermediate states in the automatic differentiation computation when applying the chain rule. For combustion flow problems, which involve short time steps and high iteration counts, the memory usage due to the chain rule can quickly become unacceptable as the complexity of the problem increases, and the speed of automatic differentiation can also significantly decrease.

To better integrate the solver's time integration process with the gradient descent optimization algorithm, JANC employs the adjoint optimization method for ordinary differential equations (ODEs) [24]. This method solves the $\partial\mathcal{L}/\partial\boldsymbol{\theta}$ by performing a backward time integration of the adjoint differential equation $\boldsymbol{N}^*$ from $t$ to $t_0$:

$$\begin{gathered}\frac{\partial \boldsymbol{a}}{\partial t}=\boldsymbol{N}^*(\boldsymbol{a}),\\ \boldsymbol{a}(t_0)=\int_{t_0}^{t}\boldsymbol{N}^*(\boldsymbol{a}(\tau))d\tau+\frac{\partial \mathcal{L}}{\partial \boldsymbol{U}(t)},\\ \frac{\partial \mathcal{L}}{\partial \boldsymbol{\theta}}=\boldsymbol{a}(t_0)^T\frac{\partial \boldsymbol{N}}{\partial \boldsymbol{\theta}}\end{gathered} \tag{6.5}$$

By solving equation (6.5) using the same TVD-RK3 numerical integration method, the gradient of the cost function can be computed with a computational load and time similar to that of the forward calculation of $\boldsymbol{U}(t;\boldsymbol{\theta})$ The specific form of the adjoint equation $\boldsymbol{N}^*$ and $\partial \boldsymbol{N}/\partial \boldsymbol{\theta}$ can be directly obtained through JAX's built-in automatic differentiation, without the need for complicated symbolic derivations. Additionally, in JAX, the JIT-compiled TVD-RK3 solver exhibits extremely fast solving speeds. When $N_t$ is large, solving the adjoint equation gradient in this way offers significant advantages over the default automatic differentiation method, both in terms of memory usage and computational time.

To implement this adjoint equation gradient computation, JANC uses the `jax.custom_vjp` function to replace the default JAX automatic differentiation, packaging the adjoint equation algorithm into a function called `make_diffeq`. This function takes the right-hand side of the differential equation $\boldsymbol{N}$ (which can be the JANC CFD solver, or any other equation such as neural network) as input and outputs a function: `ode_solver`, that can automatically solve the adjoint equation. When this function is called, `ode_solver` provides the integration results for equation (6.1) as (6.2). When the cost function containing the integration result (6.2) is invoked by JAX's automatic differentiation program, the solution of the adjoint equation is automatically carried out, efficiently and rapidly solving for the gradient.

Through `make_diffeq` and the CFD solver set up in Section 4, JANC enables the differentiability of the entire combustion flow solution process (especially the time integration process).

### *6.2 Example: Inferring injection equivalence ratio from flow field of a rotating detonation combustor (RDC)*

This section presents an example of solving the inverse problem for the flow field in a rotating detonation combustor (RDC), demonstrating JANC's adjoint optimization functionality and implementation. The RDC is an efficient propulsion device that generates thrust through the consumption of propellant by detonation waves moving around the circular combustor [33]. The speed of the detonation waves can often reach Mach 5, making the RDC flow field a typical example of complex and intense combustion flow. Furthermore, due to the periodic motion of the detonation wave around the circumference, the RDC flow field is highly transient, which makes the associated optimization tasks more challenging compared to steady-state flow fields. This section uses such a combustion flow problem to showcase the flow field complexity and solution efficiency that JANC's adjoint optimization functionality can handle.

The 2D computation of the rotating detonation flow field is carried out by unfolding the flow field of the annular combustor along the circumference and applying periodic boundary conditions on both sides. Thus, the entire computational domain is similar to the detonation tube problem, using a rectangular Cartesian coordinate system. In this section, the $x$- and $y$-directions correspond to the circumferential and axial directions of the combustor, with their lengths set as $L_x$ = 0.10m and $L_y$ = 0.04m, respectively. The computational grid is set to 1000 and 400 cells along the $x$- and $y$-axes. The upper boundary of the computational domain uses a traditional pressure outlet boundary, with a backpressure set to 0.1 atm. The lower boundary uses a premixed injection model [34], with a functional relationship between the inlet pressure and the combustor pressure to simulate the feedback relationship between combustor pressure and propellant injection. In the injection function, parameters such as the ratio of the injection area to the combustor cross-sectional area, the equivalence ratio of the premixed propellant, the total pressure, and the total temperature of the injection can be set. In this case, the area ratio is fixed at 0.2, and the total pressure and total temperature are fixed at 10 atm and 300 K, respectively. The equivalence ratio $\phi$ of the propellant is set as the parameter to be solved in this inverse problem. For simplicity, the specific heat ratio is fixed at 1.29, and the chemical reaction is modeled as a single-step hydrogen-air total combustion reaction. Detailed parameter settings can be found in the relevant example on GitHub.

The specific inverse problem solved in this section is as follows: First, set $\phi$ = 1.0 and select the flow fields $\boldsymbol{U}_0$ and $\boldsymbol{U}_{100}$ at the 20000th and 20100th time steps after ignition, as shown in Figure 11 (only temperature is shown in the figure). The inverse problem requires solving for the propellant injection equivalence ratio based only on flow field snapshots at two specific time steps.

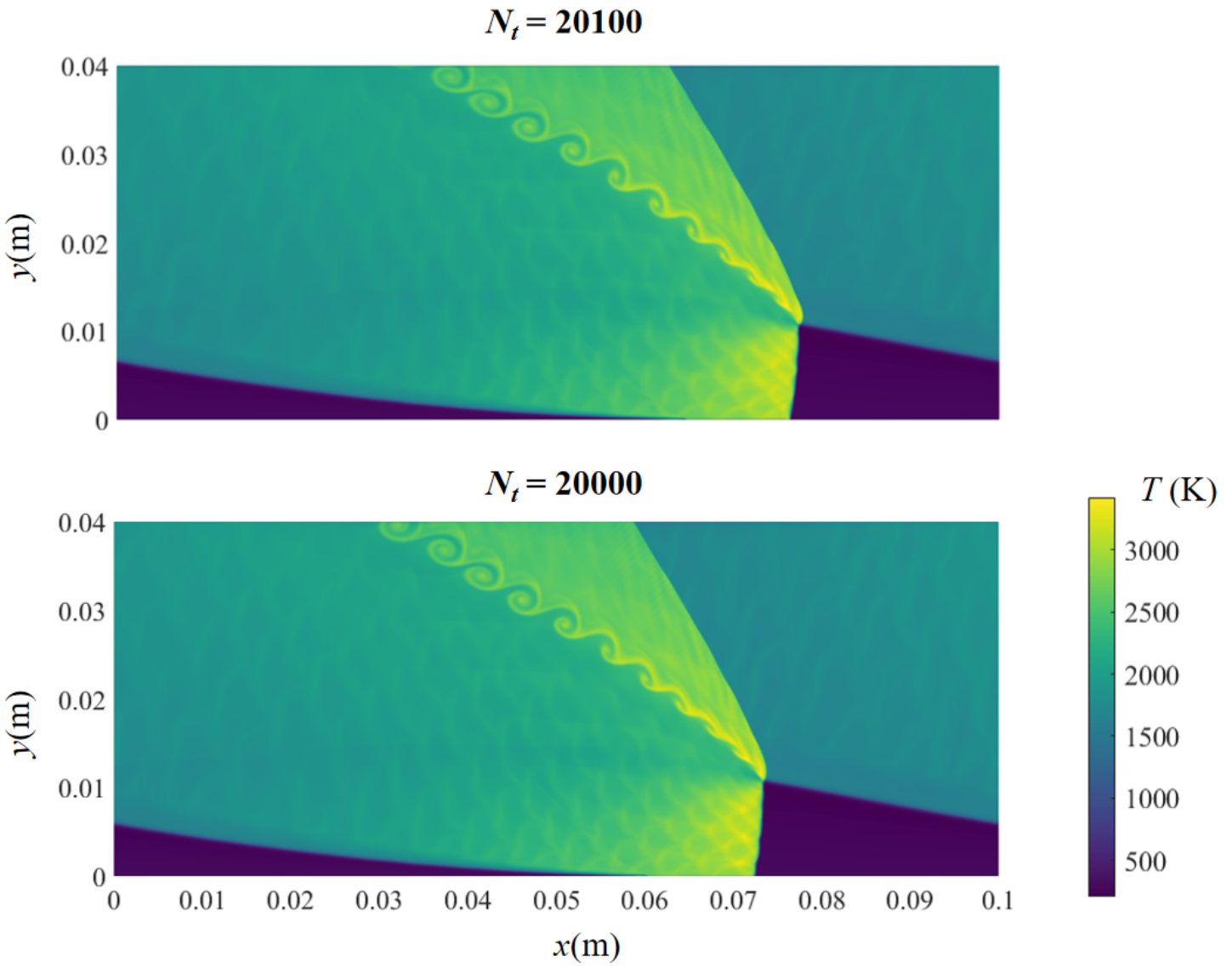

**Figure 11**. Temperature field snapshots at time steps 20000 and 20100 after ignition for equivalence ratio $\phi$ = 1.0.

To solve this inverse problem, $\boldsymbol{N}$ in equation (6.1) is directly chosen as the `rhs` function from **cfd_solver.py**, and $\boldsymbol{\theta}$ is selected as $\phi$. The initial condition for the integral in equation (6.2) is chosen as $\boldsymbol{U}_0$ The cost function is defined as the relative error norm between the flow field after 100 time steps of TVD-RK3 integration $\bar{\boldsymbol{U}}_{100}$ with parameter $\phi = \theta$ and $\boldsymbol{U}_{100}$ with $\phi$ = 1.0:

$$\mathcal{L} = \left\| \bar{\boldsymbol{U}}_{100}\left(\phi = \theta\right) - \boldsymbol{U}_{100}\left(\phi = 1.0\right) \right\| / \left\| \boldsymbol{U}_{100}\left(\phi = 1.0\right) \right\| \tag{6.6}$$

The optimization of the cost function $\mathcal{L}$ is performed using the Adams optimizer, with a learning rate set to 0.001. The initial parameters $\theta_0$ are set to 0.70, 0.80, 1.20, and 1.30 for the four different cases. The optimization results are shown in Figure 12. In Figure 12a, the value of the cost function $\mathcal{L}$ is plotted as a function of the number of iterations, while in Figure 12b, the evolution of the parameter $\theta$ over iterations is shown. It shows that under all four initial parameter sets, the parameter $\theta$ successfully converges to the correct equivalence ratio of 1.0 within no more than 400 iterations.

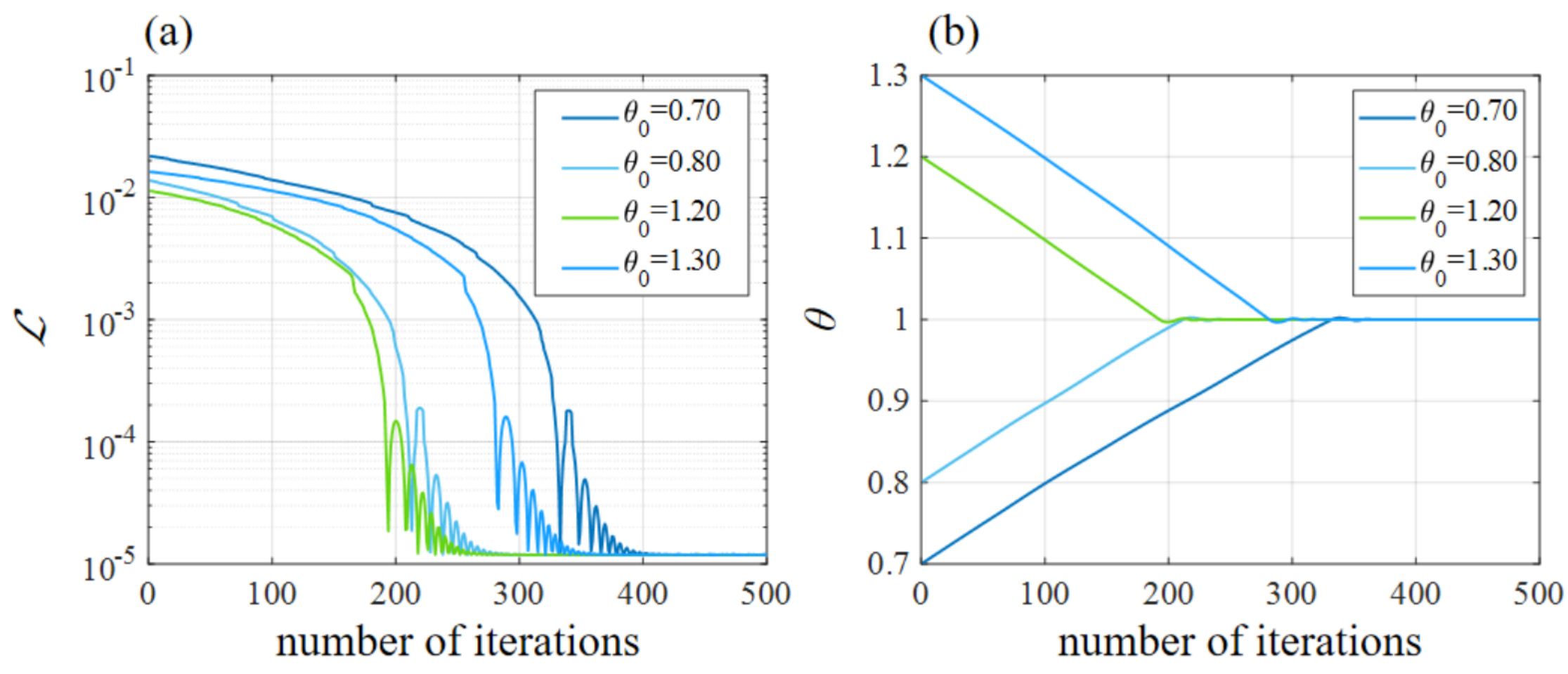

**Figure 12.** Iterative optimization under different initial parameters:(a) variation of the cost function with iteration steps; (b)evolution of the optimization parameter $\theta$ with iteration steps.

It is important to emphasize that although the RDC example is computed on 400,000 grids and each iteration requires both forward and backward propagation for 100 time steps, the adjoint optimization, compiled with JIT, achieves an average speed of 1.20 iterations of the optimizer per second on a single A100 GPU. The optimization training for a single case can be completed in 8 to 9 minutes.

For comparison, in the same inverse problem solution, the adjoint optimization is removed, and the gradient computation during training is performed using JAX's default automatic differentiation. During the first training step, JAX's XLA reports a memory overflow error:

“`XlaRuntimeError: RESOURCE_EXHAUSTED: Out of memory while trying to allocate 421565057912 bytes.`”

As can be seen, even though the A100 GPU has 40GB of memory, it is still unable to handle the memory requirements for automatic differentiation using the chain rule for the RDC optimization example with 400,000 grids and 100 time steps. This indicates that differentiable optimization of complex combustion flow problems is extremely challenging using conventional automatic differentiation methods.

## 7. Conclusions and prospects

We have developed JAX-AMR, a JAX-based adaptive mesh refinement framework. Building upon this foundation, we have created and integrated a fully-differentiable compressible reacting flows solver, which has been released as an open-source package named JANC.

JAX-AMR employs a multi-level block structure with fixed positions and shapes, enabling full compatibility with JIT compilation and vectorized operations. Currently, only the two-dimensional version is supported. The core solver of JANC is based on a structured grid in two-dimensional Cartesian coordinates, featuring WENO5 spatial discretization, 3rd-order Runge-Kutta time integration, a point-implicit chemical source advancing method, and non-dimensionalization equations. It supports fully GPU acceleration and parallel computation.

JANC and its core solver have been validated using standard benchmark cases such as the Sod shock tube and ignition delay. Parallel performance tests were conducted on a 4-GPU A100 setup. In comparative performance tests using a 2D detonation tube case against OpenFOAM, JANC's core solver demonstrated a reduction in the cost per million steps (CPMS) by two orders of magnitude. When Adaptive Mesh Refinement (AMR) is enabled in both solvers, JANC still maintains a significant advantage, with CPMS reduced to 1–2% of that of OpenFOAM. Detailed values are presented in Table 3.

**Table 3** CPMS of JANC and OpenFOAM in the detonation tube benchmark cases (Unit: USD. One A100 GPU used for JANC, 384 AMD CPUs used for OpenFOAM)

| | $CPMS_{JANC}$ | $CPMS_{OpenFOAM}$ | $CPMS_{JANC}$/ $CPMS_{OpenFOAM}$ |
|---|---|---|---|
| Core solver | 13.3 (FP64) | 1344.0 | 1.0% |
| Solver with AMR | 12.0 (synchronous)<br>6.7 (crossover) | 555.8 | 2.1%<br>1.2% |

JANC replaces JAX's default automatic differentiation with adjoint optimization, including for its time integration routines. This ensures that the entire trajectory of combustion flow dynamics is efficiently differentiable, enabling seamless integration with gradient-based optimization frameworks such as those used in machine learning. In the inverse problem example of inferring the injection equivalence ratio from RDC flow snapshots, JANC successfully identified the correct parameter under large-scale simulation conditions—400,000 spatial grid points and 100 time steps—while maintaining low memory usage and fast computation speed across multiple initial settings.

We are currently refining and enhancing the capabilities of both JAX-AMR and JANC. The ongoing development focuses on the following key enhancements:

- Extending the current two-dimensional framework to fully three-dimensional implementations;
- Integrating advanced physical models including viscous flows, turbulence model, and particle phase model to broaden application scope;
- Implementing parallel computing architecture for JAX-AMR, and optimizing the block allocation and dynamic updating algorithms;
- Developing immersed boundary methods based on JAX-AMR for complex-geometry computational domains.

These improvements are being actively developed, with plans to release new versions in the near future.

**Acknowledgement**

The authors thank Tonghui Wang from Beijing Institute of Technology for his assistance in configuring OpenFOAM benchmark cases.

## Data availability statement

JAX-AMR and JANC are now available under the MIT license at https://github.com/JA4S/JAX-AMR and https://github.com/JA4S/JANC.